\documentclass[twocolumn]{aastex63}
\usepackage{natbib,amssymb,amsmath,graphicx,here,gensymb,xcolor,txfonts,hyperref,float}

\begin{document}

\title{L-band Integral Field Spectroscopy of the HR~8799 Planetary System}

\correspondingauthor{Jordan M. Stone}
\email{jordan.stone@nrl.navy.mil}

\author[0000-0003-0695-0480]{David S. Doelman}
\altaffiliation{The first two authors contributed equally to this article}
\affiliation{Leiden Observatory, 
Leiden University, 
PO Box 9513, 
2300 RA Leiden, 
The Netherlands}
\author[0000-0003-0454-3718]{Jordan M. Stone}
\altaffiliation{The first two authors contributed equally to this article}
\affiliation{Naval Research Laboratory, 
Remote Sensing Division, 
4555 Overlook Ave SW, 
Washington, DC 20375 USA} 

\author[0000-0002-1764-2494]{Zackery W. Briesemeister}
\affiliation{Department of Astronomy and Astrophysics, 
University of California, Santa Cruz, 
1156 High St, 
Santa Cruz, CA 95064, USA}

\author{Andrew J. I. Skemer}
\affiliation{Department of Astronomy and Astrophysics, 
University of California, Santa Cruz, 
1156 High St, 
Santa Cruz, CA 95064, USA}

\author{Travis Barman}
\affiliation{Lunar and Planetary Laboratory, 
The University of Arizona, 
1629 E. Univ. Blvd., 
Tucson, AZ 85721 USA}

\author{Laci S. Brock}
\affiliation{Lunar and Planetary Laboratory, 
The University of Arizona, 
1629 E. Univ. Blvd., 
Tucson, AZ 85721 USA}

\author{Philip M. Hinz}
\affiliation{Department of Astronomy and Astrophysics, 
University of California, Santa Cruz, 
1156 High St, 
Santa Cruz, CA 95064, USA}

\author{Alexander Bohn}
\affiliation{Leiden Observatory, 
Leiden University, 
PO Box 9513, 
2300 RA Leiden, 
The Netherlands}

\author{Matthew Kenworthy}
\affiliation{Leiden Observatory, 
Leiden University, 
PO Box 9513, 
2300 RA Leiden, 
The Netherlands}

\author{Sebastiaan Y. Haffert}
\affiliation{Steward Observatory,
University of Arizona,
933 N. Cherry Ave,
Tucson, AZ 85721-0065 USA}

\author{Frans Snik}
\affiliation{Leiden Observatory, 
Leiden University, 
PO Box 9513, 
2300 RA Leiden, 
The Netherlands}

\author{Steve Ertel}
\affiliation{Steward Observatory,
University of Arizona,
933 N. Cherry Ave,
Tucson, AZ 85721-0065 USA} 

\author{Jarron M. Leisenring}
\affiliation{Steward Observatory,
University of Arizona,
933 N. Cherry Ave,
Tucson, AZ 85721-0065 USA} 

\author[0000-0001-6567-627X]{Charles E. Woodward}
\affiliation{Minnesota Institute of Astrophysics, 
University of Minnesota,
116 Church Street, SE,
Minneapolis, MN 55455, USA}

\author{Michael F. Skrutskie}
\affiliation{Department of Astronomy, University of Virginia, Charlottesville,
VA 22904, USA}

\begin{abstract} Understanding the physical processes sculpting the appearance
of young gas-giant planets is complicated by degeneracies confounding effective
temperature, surface gravity, cloudiness, and chemistry. To enable more
detailed studies, spectroscopic observations covering a wide range of
wavelengths is required.  Here we present the first L-band spectroscopic
observations of HR~8799~d and e and the first low-resolution wide bandwidth
L-band spectroscopic measurements of HR~8799~c.  These measurements were
facilitated by an upgraded LMIRCam/ALES instrument at the LBT, together with
a new apodizing phase plate coronagraph. Our data are
{generally} consistent with previous photometric
observations covering similar wavelengths, {yet there
exists some tension with narrowband photometry for HR~8799~c}.  With the
addition of our spectra, each of the three innermost observed planets in the
HR~8799 system have had their spectral energy distributions measured with
integral field spectroscopy covering $\sim0.9$ to $4.1~\mu\mathrm{m}$. We
combine these spectra with measurements from the literature and fit synthetic
model atmospheres. We demonstrate that the bolometric luminosity of the planets
is not sensitive to the choice of model atmosphere used to interpolate between
measurements and extrapolate beyond them. Combining luminosity with age and
mass constraints, we show that the predictions of evolutionary models are
narrowly peaked for effective temperature, surface gravity, and planetary
radius. By holding these parameters at their predicted values, we show that
more flexible cloud models can provide good fits to the data while being
consistent with the expectations of evolutionary models.  
\end{abstract}


\section{Introduction} 

More than a decade of direct imaging photometric and spectroscopic probes of
gas-giant exoplanets have provided an important understanding of the physical
processes sculpting their atmospheres. The HR~8799 system, which includes four
giant planets \citep{Marois2010}, is by far the most well studied system for
direct imaging. In addition to the appeal of comparing the appearance of
multiple coeval planets, HR~8799 is also observable from both hemispheres,
includes a bright host star required for high-performance adaptive optics (AO)
systems, and the planets are observed with projected separations and contrasts
amenable for modern AO instruments at the world's largest telescopes. In fact,
the outermost planet falls outside the narrow field of view of many of the
latest high angular resolution instruments.

Early studies identified that the HR~8799 planets occupied a rarefied locus of
near-IR color magnitude diagrams \citep{Marois2008}, being redder and/or
fainter than typical brown dwarfs with similar effective temperatures. Model
atmosphere fits to the HR~8799 planets, and to other young directly imaged
planetary mass companions, match these measurements reasonably well
\citep{Patience2010} but with scaling factors that implied planet radii
($<1~R_\mathrm{Jup}$) which are much too small to be consistent with our
understanding of gas-giant planetary structure.

Atmospheric modelers and brown dwarf observers quickly aided our understanding
of some of these observations by pointing out that atmospheres, especially
substellar atmospheres, are not single parameter systems described only by
effective temperature. Surface gravity, particularly for young planets that are
low mass with extended radii, is an essential consideration for
a proper interpretation of the data \citep[see][]{Stephens2009, Barman2011a,
Marley2012}. Low gravity atmospheres can loft clouds above their photospheres
at cooler temperatures than higher gravity objects \citep{Barman2011a,
Marley2012}. Additionally, low-gravity atmospheres are more susceptible to
vigorous mixing that can alter the balance of chemical species in the
photosphere, including the relative abundance of methane and carbon monoxide
\citep{Hubeny2007}.

Even so, model fits to data are plagued by degeneracies between temperature,
gravity, cloudiness, and chemistry \citep[see][]{Currie2014}. Cloud structure
in particular is confounding because of the complex physics governing cloud
formation (and dissipation) and because of the number of parameters needed to
describe them, including cloud thickness, cloud coverage (patchy/homogeneous),
cloud particle size distribution, and cloud composition, among others.
{For the HR~8799 planets in particular, models with
either homogeneous cloud coverage and small grain size
\citep[e.g.,][]{Bonnefoy2016, Konopacky2013, Greenbaum2018} or patchy cloud
models \citep[e.g.,][]{Currie2011, Skemer2014, Currie2014} can provide
reasonable fits to the data.}

Detailed narrowband spectroscopic observations can enable studies to
characterize certain aspects of planetary atmospheres in ways that do not seem
to depend on the details of cloud structure, such as C/O ratio
\citep{Konopacky2013, Barman2015, Wang2020, Molliere2020}. Photometric studies constraining
a large portion of the planetary spectral energy distribution (SED) can be
successful in breaking model degeneracies to constrain planet composition
\citep{Skemer2016}.
Notably, \citet{Wang2020} find that their free retrieval with L- and M-band data yields solutions that are closer to physically and chemically motivated models compared to excluding this wavelength range,  and remark that this data data helps constrain the abundances and cloud condition.
The Arizona Lenslets for Exoplanet Spectroscopy
\citep[ALES,][]{Skemer2015,Skemer2018} instrument was built to increase the
wavelength coverage of high-contrast spectroscopic observations to improve our
understanding of gas-giant atmospheres.

In this paper, we present the first L-band spectroscopy of HR~8799~d, and e.
For HR~8799~c {some previous spectroscopic
observations exist at these wavelengths, including the early work of
\citet{Janson2010} presenting three spectral channels covering a small range of
the atmospheric window, and the high-resolution work presented by
\citet{Wang2018}. We present the first broadband low-resolution spectroscopy of
HR~8799~c in the L-band.}

After describing our observations and data reduction approach in Section
\ref{sec:observations_and_data_reduction}, we compare our measurements to those
from the literature in Section \ref{sec:Comparison} finding general
agreement with earlier photometric measurements, although we identify some
tension at LNB5 and LNB6 for HR~8799~c. With the addition of our data, each of
the three innermost directly imaged planets in the HR~8799 system have had
their emission spectra measured with low-resolution integral field spectrograph
(IFS) spectroscopy spanning $\sim0.9-4.1~\mu\mathrm{m}$.  In Section
\ref{sec:modelfitting} we compile data from the literature for each planet and
describe a model fitting approach to match spectra from two families of
synthetic atmosphere models and blackbodies.  The results of our fitting are
presented in Section \ref{sec:Results}. Our initial fitting approach did not
impose any restrictions on planet radius or other bulk quantities. We show that
the Barman/Brock family of models \citep{Barman2011a, Brock2021} are capable of
providing reasonable fits to the data as well as reasonable planet radii in
some cases, but that the radii required for the DRIFT-Phoenix models
\citep{Witte2011} were not consistent with expectations based on evolutionary
models of gas-giant structure. As expected, the blackbody models provided
neither a good approximation to the data nor reasonable radii.

{Many previous studies have appealed to evolutionary
models to constrain their atmospheric modeling efforts
\citep[e.g.,][]{Barman2011a, Marley2012, Konopacky2013, Rajan2017, Brock2021}.
In Section \ref{sec:discussion} we develop a Monte Carlo approach to generating
quantitative priors for atmospheric model fitting. This approach incorporates
the details of constraints on system parameters such as age, mass, and
luminosity, and results in priors for $T_{\mathrm{eff}}$, $\log(g)$, and radius
that can be directly tied to specific evolutionary models.}

{For the HR~8799 planets, we point out that the
luminosity of each, with such broad spectroscopic coverage, is tightly
constrained ---depending little on the choice of well-scaled atmospheric model
used to interpolate between observations and extrapolate beyond them. We use
this luminosity together with constraints on system age and planet masses, to
show that the predictions of hot-start luminosity models are narrowly peaked in
effective temperature, surface gravity, and radius.} We follow the example of
\citet{Brock2021} and rerun our fits, fixing effective temperature, surface
gravity and radius, and using more flexible atmospheric models to explore what
can be inferred about cloud structure assuming gas-giant evolution models are
reliable. Finally, in Section \ref{sec:conclusion} we summarize our results and
comment on future applications of both the technology demonstrated and the
analysis performed particularly towards Gaia-detected companions.
    
\section{Observations and data reduction}\label{sec:observations_and_data_reduction} 

We observed HR~8799 on 2019 September 18 for 1 hour 53 minutes with the
upgraded LBTI/ALES instrument \citep{Skemer2018, Hinz2018}.  ALES is an
adaptive optics-fed integral field spectrograph with sensitivity out to
5 microns \citep{Skemer2015,Skemer2018} and is used as a mode of the LMIRCam
  instrument \citep{Skrutskie2010, Leisenring2012}, part of the Large Binocular
Telescope Interferometer architecture \citep{Hinz2016}. We used the main mode
of ALES with a square field of view of $\sim$2.2 arcseconds on a side with
a spectral resolution of $\sim$35, spanning the $2.8-4.2$ $\mu$m range. The
detector integration time was set to 3.934 seconds and the first and last
0.492-second read of each ramp was saved to enable the subtraction of detector
  reset noise (correlated double sampling).  The conditions were stable with
seeing between 0.8 and 1.1 arcseconds. We acquired 1300 frames on-target for
a total of 1 hour 24 minutes.  The LBTI architecture does not include an
instrument derotator and our images include a total field rotation of
$85\fdg64$ through meridian crossing.  

The observations were conducted as part of early characterization efforts using
a new apodizing phase plate upstream of the IFS within LMIRCam.  The
double-grating 360$^{\circ}$ vector apodizing phase plate
\citep[dgvAPP360][]{Doelman2020, Wagner2020} suppresses the stellar diffraction
halo by multiple orders of magnitude over the full $2-5~\mu\mathrm{m}$
bandwidth. The dgvAPP360 is different from the more commonly used grating-vAPP
\citep{Snik2012, Otten2017,Doelman:21}, which creates two images of the star
each with a D-shaped dark zones on opposite sides. The additional grating in
the dgvAPP360 diffracts the light back on-axis, such that the two apodized
images overlap, resulting in a single image of the star.  Furthermore, the
phase design of the dgvAPP360 creates a dark zone in a full annulus (covering
360$^{\circ}$) surrounding the star. The resulting point spread function (PSF)
is much smaller compared to the gvAPP and is consequently better suited for the
small field of view of an IFS.  

As a pupil plane optic, the dgvAPP360 is particularly well suited for ALES, as
careful alignment of the IFS magnifiers with a focal plane spot is not
necessary. {Since the dgvAPP360 response is tip/tilt
invariant, drifts in the PSF location during observing do not result in a loss
of performance, i.e. dark zone contrast. Additionally, the location of the star
is known in every frame. This is ideal because with our short thermal-IR
exposure times we are able to increase contrast with post-processing shift and
add techniques.}

In order to increase on-source efficiency we chose not to periodically nod to
a sky position to track variable background emission. Instead, we collected
a total of 99 background frames, where the first 13 were taken after 100
science frames and the other 86 directly after the science sequence.  We
achieved a ratio of on-target to background integration of 93\%.  However, as
described {in Section \ref{sec:classical_adi}}, our
original plan for removing the sky background at each wavelength within each
cube was complicated as a result of an instrument related issue, requiring
a more sophisticated data reduction approach than envisioned at the time our
observations were designed. {This issue is the movement
of the stellar PSF and associated structures with respect to non-uniform
thermal background. }

Wavelength calibration of our spectral cubes was achieved by observing through
four narrowband ($R\sim100$) filters. The filters, spanning $2.9~\mu\mathrm{m}$
to $3.9~\mu\mathrm{m}$, are all located upstream of the ALES optics within
LMIRCam, and are observed sequentially \citep{Stone2018}. Thermal emission from
the sky fills the ALES field of view and provides fiducial wavelength spots at
every position.  Images were saved with 3.934 second exposures. For the two
shorter wavelength filters we saved 25 frames each. For the two longer
wavelength filters we saved 10 frames each. We saved 200 dark frames with the
same exposure time.

\subsection{Raw Frame Preprocessing}

Prior to making cubes, each of our ALES frames were preprocessed to correct for
reset noise, variable channel offsets, hot pixels, and a fixed light leak from
within the instrument that causes off-axis light to pass through the lenslet
array and fill some of the pixels between the on-sky spectral traces. This
light leak can bias the measured position of the wavelength calibration spots
and result in an inaccurate estimate of some of the spectral spatial profiles,
significantly affecting the quality of our spectral cubes.  

For each image we subtracted the first read from the last. This removes the
reset noise and most of the channel offsets seen in the raw images. Residual
channel offsets were then removed using a median of the $4\times64$ lower
overscan pixels in each channel. We noticed that the 127th and 128th columns
(and the corresponding columns every 128 pixels) behaved differently than the
other columns within their channels, so we treated these individually,
subtracting only the median of the lower four overscan pixels in the same
column. In the orthogonal direction, eight overscan pixels in each row are
median combined and the resulting 2048 pixel column is then smoothed with
a Savitzky-Golay filter using a window length of 31 and polynomial order of 3.
The resulting smoothed overscan column is then removed from each column in the
image.

\begin{figure*} 
\centering 
\includegraphics[width = 0.9\linewidth]{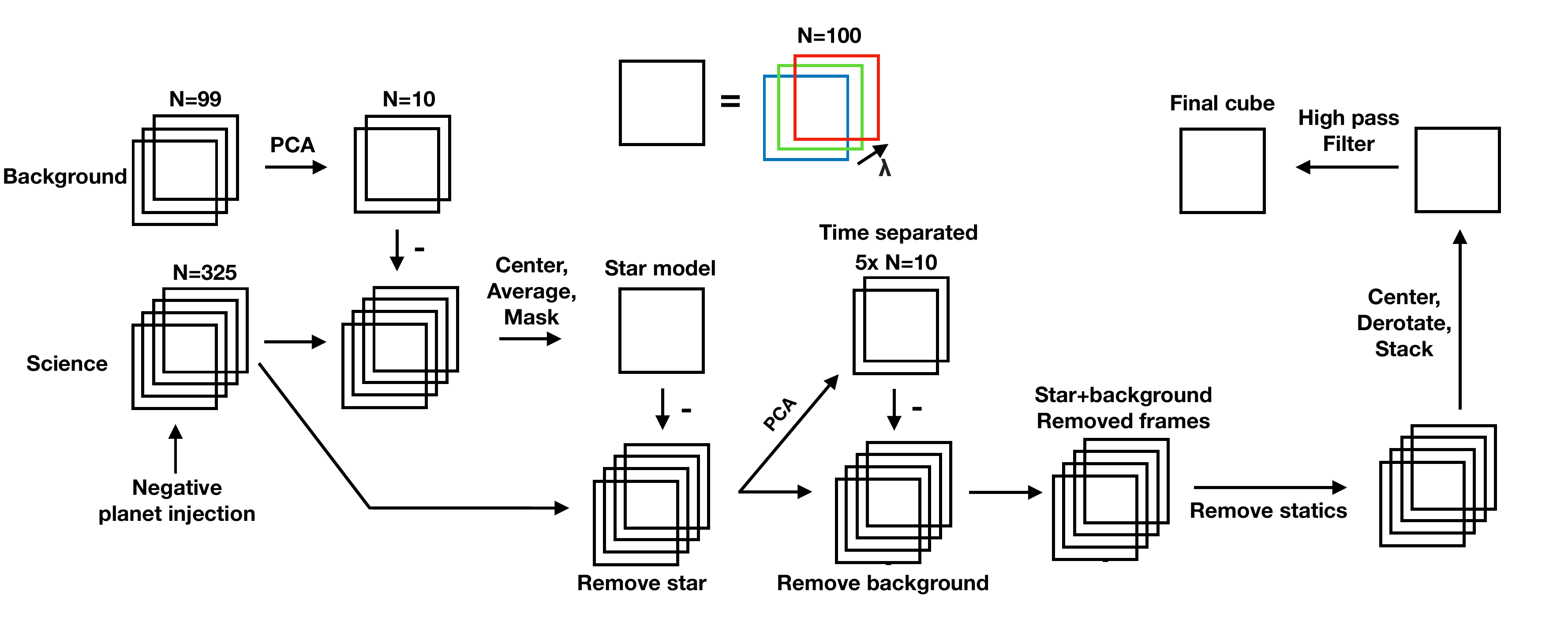} 
\caption{Flowchart of the data-reduction method.} 
\label{fig:flowchart_data_red} 
\end{figure*}

To correct for the light leak, an empirical model of the leak was subtracted
from each frame. To build this model, we first median combined the narrowband
filter wavelength calibration images for each of the four filters. The
resulting medians were then median combined. This approach removes the narrow
spots, leaving behind only the light leak signal.  

Bad pixels in each processed frame were replaced using the median of the
4 nearest good pixels. Bad pixels were identified as overly hot in dark frames
and/or overly cold in flat illuminated frames.

\subsection{Spectral cube extraction} 

We extracted $(x, y, \lambda)$ data cubes using an inverse variance and spatial
profile weighted extraction approach on each of the $63\times67$ micro-spectra
across the ALES field \citep{Horne1986, Briesemeister2018}. To build extraction
weights we used the 99 sky images to define the spatial profile and variance.
We built the spatial profile for each microspectrum assuming a constant profile
with wavelength and median combining along wavelength direction. To mitigate
crosstalk we enforced a seven pixel wide window, which accommodates the full
width at half maximum for spectra near the center of the field of view, but
crops more light for some aberrated spectra near the edge of the field of view.
For each of the microspectra we also masked out the right side of the spatial
profile for the bluest wavelengths of the spectrum where the risk of
contamination (spectral crosstalk) from the brightest red part of the
neighboring microspectrum is highest.

A quadratic wavelength solution, {mapping pixel
position to wavelength}, was fit to each spectrum using the peak pixel for
each narrowband wavelength filter image and the corresponding wavelength from
a cryogenic filter trace. {Since each of the
microspectra are not sampled in exactly the same way by the pixels of LMIRCam,
in order to produce a spectral cube with constant wavelength at each slice},
cubic interpolation was used on each spectrum to extract the same wavelengths
at each position. 

The new ALES lenslet array has lower amplitude optical aberrations than the
previous array, but the spot produced by each lenslet is affected by a residual
astigmatism whose axis rotates as a function of position in the array. This
creates a varying spatial profile and a varying spectral resolution that both
contribute to a varying throughput as a function of position and wavelength.
A lenslet flat was generated to quantify this throughput by extracting a cube of
the median sky image and normalizing each wavelength slice. This flat is then
used to correct each of the science cubes.

As a final step, we binned the data per four frames in time by averaging,
reducing the number of cubes from 1300 to 325.

\subsection{High Contrast Image Processing} \label{sec:classical_adi} 

{As mentioned before, our original plan for removing
the sky background at each wavelength within each cube is complicated by the
issue of movement of the stellar PSF with respect to a non-uniform thermal
background.  The thermal background has spatial structures, of which the total
intensity varies in time, yet the relative intensity of the structures are
constant.  During the observation sequence the PSF moves with respect to these
background structures in a u-shape.}  This u-shape is $\sim35$ mas (=1 spaxel)
in the x-direction and $\sim70$ mas in the y-direction, while the
frame-to-frame jitter is $\sim4$ mas.  A known source of PSF movement in ALES
is the lenslet array, which moves due to flexure of the instrument with
telescope pointing.  The movement of the PSF on the detector is correlated with
elevation, suggesting that flexing is indeed a contributor.
{A possible second contributor is the atmospheric
dispersion separating the visible star that remains fixed by the AO, and the
Thermal-IR star which will move along the position of the star in the
y direction.}
\begin{figure}
    \centering
    \includegraphics[width = \linewidth]{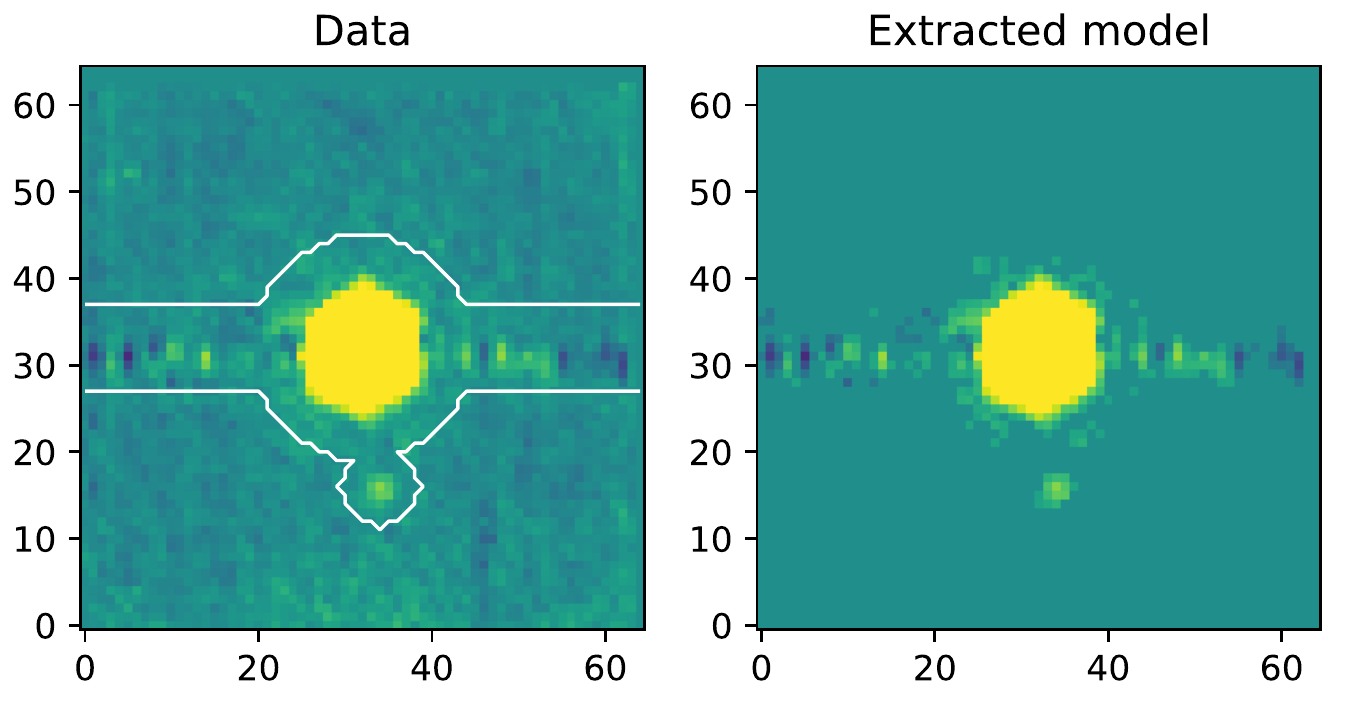}
    \caption{\textit{Left:} Background-subtracted PSF at 3.55 $\mu$m, using 10
    PCA components of the 99 background frames. Residual structures indicate
    the presence of a ghost, some speckles, and a horizontal periodic
    structure. \textit{Right:} Same as left, but clipped at $2\sigma$ inside
    the area of a mask, indicated by the white lines in the left-hand image. }
    \label{fig:PSF_model_static_bg}
\end{figure}

{The total PSF motion with respect the thermal
background structures complicate the data reduction.  This decoupled motion is
difficult for standard angular differential imaging \cite{Marois2006}
processing approaches that center on the star and results in non-optimal
removal of the thermal background.  As we chose to not frequently nod to sky we
have a limited number of background frames.  Using only these 99 frames will
give background-subtracted science frames that are limited by photon noise
related to the thermal background.  The solution is to use science images to
calculate the thermal background, of which there are 1300 before binning.
Extracting a model of the thermal background from the science image is not
straightforward.} In addition to the thermal background, the science frames
also contain the stellar and planetary PSFs.  Because the planets move across
the detector with the parallactic angle we can mask their predicted locations,
and generate an estimate of the background. The star constitutes a larger
problem, as possible quasi-static speckles or ghosts will contaminate the
estimated background.  In addition, the background on the stellar PSF location
is inaccessible, which makes accurate stellar photometry impossible.
\begin{figure}
    \centering
    \includegraphics[width = 0.90\linewidth]{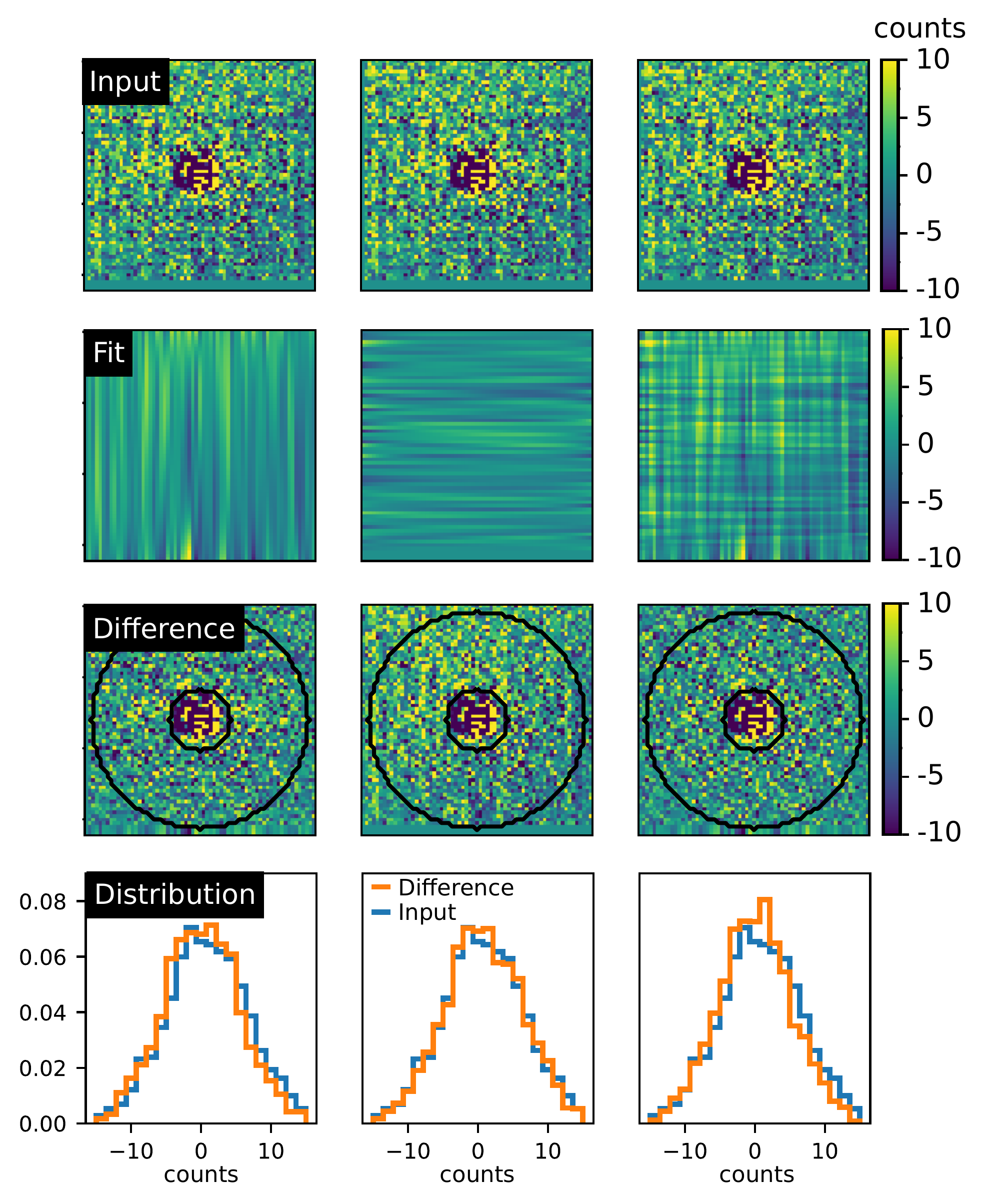}
    \caption{Removal of row and column discontinuities by fitting a third-order
polynomial to every row and column.}
    \label{fig:structured_noise}
\end{figure}

We introduce a combined method where we model the star and background
separately and subtract them from all frames for each wavelength.  A flowchart
of the method is shown in Fig. \ref{fig:flowchart_data_red}.  We start by
subtracting the 99 background frames from the science data to create a stellar
PSF model.  {The additional background noise from the
subtraction is much lower than the local stellar photon noise.} Therefore, an
accurate model of the stellar PSF can be extracted by centering and co-adding
the background subtracted frames.  {The combined images
revealed additional structures that are co-moving with the star on the
detector.  These are an optical ghost arising from the L-band filter and
electrical ghosts caused by inter-channel capacitive coupling on the detector
\citep{Finger2008}.  They are shown in in Fig. \ref{fig:PSF_model_static_bg},
in addition to our method to extract these features for the stellar model.  We
create a mask that surrounds these features for every wavelength bin and only
keep the signal that is more than $2\sigma$ compared to the background outside
of the mask. } Removing the stellar PSF from the science data using this PSF
model is now straightforward.  We fit a decentered PSF model to background
subtracted data frames for each wavelength, retrieving the stellar PSF
intensity and the science frames without the stellar PSF.

After masking the planets, we can model the background from the star-subtracted
science frames. For every frame we select the frames separated in time by 30
minutes to minimize self-subtraction.  {We note that
the change of the parallactic angle with time during the observations is
relatively constant, with 24, 30, and 24 degrees change in three 30 minute
windows.   From the time separated frames we calculate 10 PCA components.} We
optimally subtract these components, such that the residuals in the frame are
dominated by the photon noise of the star and background.
{This method of background subtraction} results in
cubes with a subtracted star and background, where the background is calculated
from the science frames themselves. Now, we can check if this method indeed
reduces the background noise compared to simply using 10 PCA components of the
99 background frames.  By subtracting 10 PCA components from the raw data for
both methods, we compare the standard deviation of the residuals after masking
the star.  Between $2.9$ $\mu$m and $4.2$ $\mu$m, we obtain a reduction of 10\%
in the standard deviation of the residuals using the background estimate from
the on-source frames compared to the background model derived from the
off-source frames.  {Therefore, we use the science data
background estimate for all further data reduction.}
\begin{figure}
    \centering
    \includegraphics[width = 0.95\linewidth]{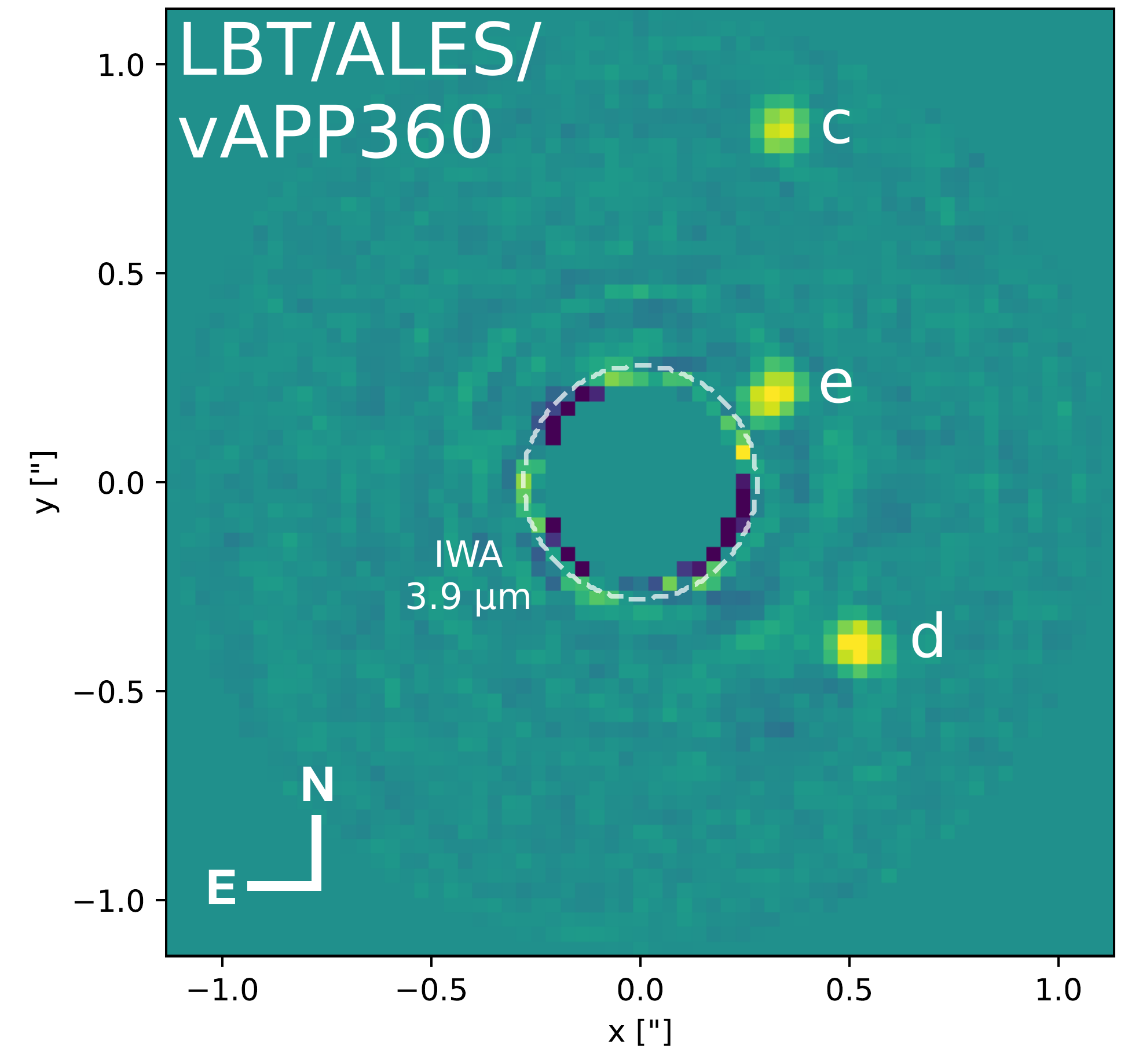}
    \caption{LMIRCam/ALES image of HR~8799~c, d, and e using the dgvAPP360 coronagraph. The
    final image is the combination of the individual wavelength slices between
    3.55-4.15 $\mu$m. North is up, East is left.}
    \label{fig:HR8799_detection}
\end{figure}
We inspect the star- and background-subtracted frames for residual structure by
averaging them in time and wavelength.  The residuals are not well described by
purely Gaussian noise, but contain structures that are column and row specific
and vary in time, see Fig. \ref{fig:structured_noise}.  The structures are
faint (1-10 counts) and {originate from the way in which the ALES
microspectra intersect the different channels of the LMIRCam detector}. These
discontinuities are characterized by fitting polynomials of the third order to
each row and column, which are shown in Fig. \ref{fig:structured_noise}.
A third order polynomial is low-order enough over the 65 pixels that it is
minimally affected by planet signal, but to be sure we mask the star with an 18
pixel circular mask and the planets with a 5 pixel circular mask and remove
those pixels from the fit.  Using a Kolmogorov-Smirnov test we verify that both
the before and after distributions are non-Gaussian, however we find that the
average of the noise distribution is now consistent with zero and the standard
deviation of a background region is reduced by 10\%.  We note that the
polynomial background fit has a large number of variables for the full image,
but we found it to be the only method that captured the behavior of this
phenomenon.  Combined with the stellar PSF removal and the background removal,
the row and column fits remove most structures present in the data in a way
that minimizes self-subtraction of planets.     

We center and derotate the background and star subtracted cubes and median
stack them along the time axis to create a single master image cube. The final
cube is put through a high-pass filter where we remove global structures on the
background by subtracting a Gaussian smoothed frame for each wavelength in the
final cube.  The Gaussian has a standard deviation of 5 pixels (FWHM = 11.8
pixels) and {we mask the locations of the HR~8799
planets and the star with NaN values.  These NaN values were interpolated over
using Astropy.convolve  \citep{astropy2013} to retrieve an estimate of global
structures in the background inside the mask.}

\subsection{{Final sensitivity} }

\begin{figure}
    \centering
    \includegraphics[width = 0.95\linewidth]{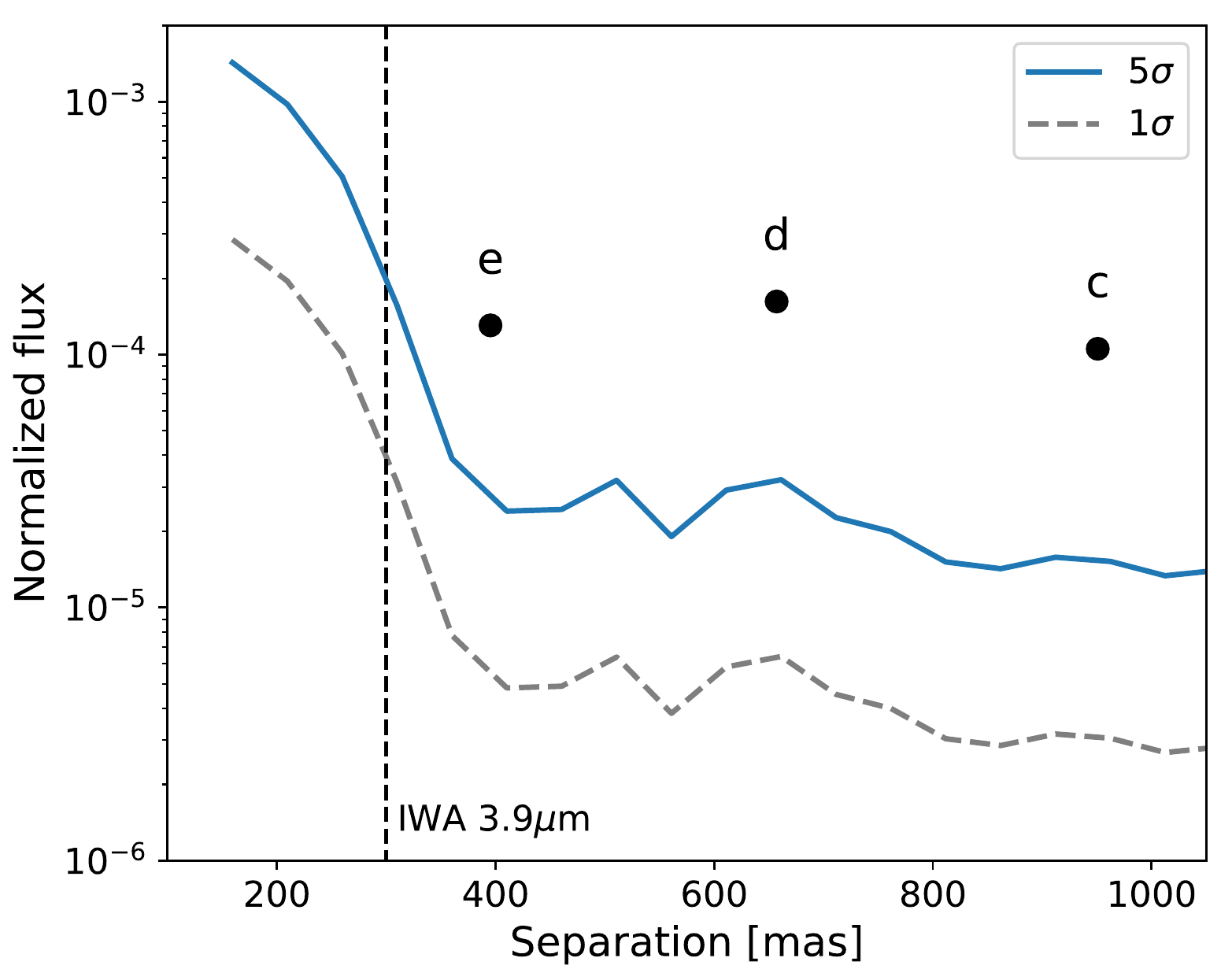}
    \caption{Contrast as a function of separation using LMIRCam/ALES with the
    dgvAPP360. The background limit is quickly reached outside the inner working
    angle of the coronagraph.}
    \label{fig:HR8799_contrast_curve}
\end{figure}

HR~8799~c, d, and e are detected with high S/N using the combination of ALES
with the dgvAPP360.  At the location of HR~8799~e there are some residuals from
speckles, yet the residuals at the locations of HR~8799~c and d are dominated
by the thermal background.  HR~8799 b is outside of the field of view of ALES.
To estimate our sensitivity we create a crude contrast curve using our data. We
focus on the wavelengths where the throughput of the dgvAPP360 is highest,
combining images between $3.5$ $\mu$m and $4.15$ $\mu$m.  The wavelength-binned
image is shown in Fig.  \ref{fig:HR8799_detection}, where all three planets are
clearly visible. At each radius, a ring of subapertures is created, each with
diameter of 1.7$\lambda/D$, avoiding the planets when necessary.  Flux within
each subaperture is summed and the standard deviation of fluxes at each radius
is taken as the noise. Contrast is determined by performing similar aperture
photometry on the primary star. The resulting contrast curve, not corrected for
varying sample size with radius, is shown in Fig.
\ref{fig:HR8799_contrast_curve}, and quickly reaches a noise floor beyond the
inner working angle.  HR~8799~c, d, and e are detected with S/N ratios of 29,
25 and 19 respectively. 

\subsection{{Spectral extraction}} \label{sec:spectralExtraction}

To extract contrast spectra for each planet from the data, we inject negative
planet signals at the locations of the HR~8799 planets in each frame.  The
injected planets are scaled copies of the stellar PSF at each wavelength. This
is a unique strength of the dgvAPP360: we have an unsaturated stellar PSF that
acts as a reference PSF for the planets for every science exposure.  The
planet-subtracted cubes are reduced using the same method described above and
in Fig.  \ref{fig:flowchart_data_red}.  For each planet and each wavelength bin
we optimize the planet location and amplitude by evaluating the Hessian at the
planet location in a circular aperture with a diameter of 7 pixels. The Hessian
is a measure of the curvature of the image surface, which is minimal when the
planet is completely removed \citep{Stolker2019}.  { Additionally, we
find that the distribution of the retrieved locations has a standard deviation
of 0.4 pixels for HR 8799 c and d, and 0.6 pixels in HR 8799 e.  The resulting
flux loss for a PSF that is masked with a circular aperture of 7 pixels with
a shift of 0.6 pixels is around 8\%.  This is smaller than the error bars in
the retrieved flux calculated from bootstrapping.}

To generate error bars for the retrieved contrast spectra we apply
bootstrapping to the data reduction method, selecting 325 frames (with
replacement) at random from the data 50 times in total.  For the purposes of
the bootstrap, the fluxes of the planets for each wavelength are retrieved
using aperture photometry, rather than fake planet injection, as the data
reduction method is computationally expensive.  Assuming that the
self-subtraction is similar for all iterations, the distribution of retrieved
amplitudes should be a good approximation of the distribution with negative
planet injection.  {The standard deviation of
measured fluxes at each wavelength is the $1\sigma$ error due to random noise.
Bootstrapping cannot be used to estimate systematic or persistent issues with
the data. The scatter seen in the spectrum HR~8799~e, may indicate that the
data are influenced by residual speckle noise, especially toward shorter
wavelengths.}

  \begin{figure*}
    \centering
    \includegraphics[width = 0.9\linewidth]{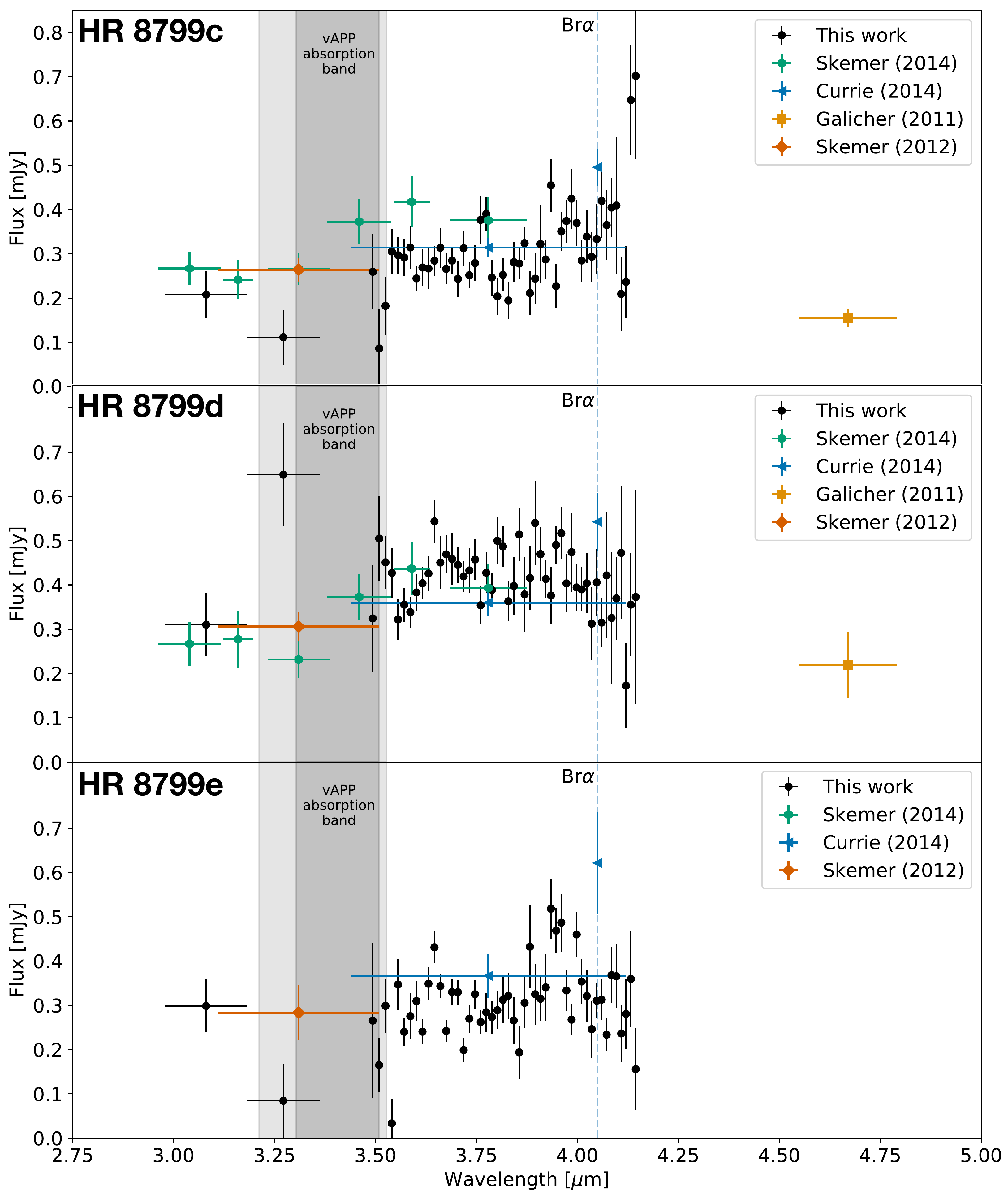}
    \caption{Spectra recovered with negative planet injection for HR~8799~c, d, and e,
    showing apparent flux. We compare our LMIRCam/ALES spectra to the 3.3 $\mu$m magnitudes from
    \cite{Skemer2012}, the narrow-band magnitudes from
    \cite{Skemer2014}, the broad $L$'-band and [4.05] magnitudes from
    \cite{Currie2014}, and the M-band magnitudes from \citep{Galicher2011}.
    The light gray and dark gray vertical swaths correspond to the vAPP absorption band, with
    an absolute transmission of 40\% and 25\% respectively, as shown in Figure 
    \ref{fig:throughput}. }
    \label{fig:final_spectra}
\end{figure*}

ALES made no significant detection of any planet between 3.35~$\mu$m and
3.5~$\mu$m due to the absorption of the dgvAPP360 coronagraph (see Appendix
\ref{sec:vAPP_characterization} for spectral characterization of the
dgvAPP360). We bin the data between 2.99 and 3.17 $\mu$m and 3.17 and
3.35~$\mu$m to retrieve two photometric points for wavelengths short of the
dgvAPP360 absorption feature.  For this purpose, the negative planet is
injected for all wavelength slices separately with a constant spectral slope
and the final evaluation of the Hessian is performed on the median combined
images.  Bootstrapping is applied to find the error on these measurements as
well. 

We perform flux calibration of the planet contrast spectra by multiplying by
a calibrated spectrum of the primary star. We used the SED analyzer VOSA
\citep{Bayo2008} to fit a {BT-Settl} model to the SED
of the host star including data from Tycho2, 2MASS, and WISE
\citep{Hog2000,Cutri2003, Cutri2012}. We retrieved a temperature of 7200 K,
log(g) = 4 log(cm/s$^2$), a metallicity of
0.5, an $\alpha$ = 0 and a multiplicative dilution factor of 6.416e-19.  We
  smoothed this BT-Settl model to the resolution of ALES and sampled it at the
same wavelengths as our final cube. We then multiply this calibrated smoothed
and sampled spectrum of the star by the contrast spectra of the planets to
yield flux calibrated spectra of HR~8799~c, d, and e.
{The retrieved spectra can be found in Appendix
\ref{App:table_spectra}}.

\subsection{Comparison to other measurements in the band}
\label{sec:Comparison}

Figure \ref{fig:final_spectra} compares our ALES measurements with previous
thermal-IR measurements in the literature for HR~8799~c, d, and e.
{To quantitatively compare our ALES measurements to
literature measurements with wider bandwidths than our spectral channels, we
calculate synthetic photometry using our ALES spectra and cryogenic filter
traces provided by the Spanish Virtual Observatory \citep{Rodrigo2012,
Rodrigo2020}. Since the NIRC2 $L^{\prime}$ band extends into the vAPP
absorption band where we do not have ALES measurements, we interpolate through
the vAPP absorption band to the $3.1~\mu\mathrm{m}$ synthetic photometry point.
This is a reasonable interpolation because \citet{Skemer2014} observed no
significant absorption through these wavelengths. We cannot synthesize
a photometric measurement for the Br-$\alpha$ narrowband filter photometry
presented by \citet{Currie2014}, or for the LNB1 and LNB2 filters presented by
\citet{Skemer2014} as these filters all have narrower bandwidths than the ALES
spectral channels.}

{We use a Monte Carlo approach to propagate
correlated uncertainty in the ALES measurements to the synthetic photometry.
First, the spectrum of each planet is modeled as a multi-dimensional Gaussian
distribution with means determined by our spectra and covariance estimated
using the method of \citet{Greco2016}.  Next, photometry is measured for each
of 100 draws from these distributions and the uncertainty taken as the standard
deviation of the measurements. Table \ref{table_photometry} lists the
results.}

{Our flux scaling and photometry for the system is
consistent with photometry presented by \citet{Currie2014}, the ALES
$L^{\prime}$ measurements appearing a bit low for planet c, a bit high for
planet d, and nearly the same for planet e. For planets c and d, we can also
compare to the LMIRCam-LNB5 and LNB6 measurements from \citet{Skemer2014}. For
planet c there exists some tension, with the ALES measurements fainter by
2.35$\sigma$ and 1.95$\sigma$ for LNB5 and LNB6, respectively. For planet d,
  the ALES measurements seem consistent with the previous results.}

\begin{deluxetable}{lccc}
\label{table_photometry}
\tablecaption{{Comparing to Photometry}}
\tablehead{\colhead{Filter} &
           \colhead{ALES synth. phot.} &
           \colhead{Lit. phot.} &
           \colhead{Difference}\\
           \colhead{} &
           \colhead{(mJy)} &
           \colhead{(mJy)} &
           \colhead{($\sigma$)\tablenotemark{$^{\dagger}$}}
           }
\startdata
\cutinhead{planet c}
NIRC2-$L^{\prime}$ & $0.286\pm0.01$ & $0.337\pm0.02$\tablenotemark{$^{a}$} & -1.8\\
LMIRCam-LNB5 & $0.273\pm0.02$ & $0.435\pm0.06$\tablenotemark{$^{b}$}  & -2.35\\
LMIRCam-LNB6 & $0.271\pm0.01$ & $0.392\pm0.05$\tablenotemark{$^{b}$}  & -1.95\\
\cutinhead{planet d}
NIRC2-$L^{\prime}$ & $0.437\pm0.02$ & $0.387\pm0.03$\tablenotemark{$^{a}$} & 1.6 \\
LMIRCam-LNB5 & $0.366\pm0.02$ & $0.436\pm0.06$\tablenotemark{$^{b}$} & -1.12 \\
LMIRCam-LNB6 & $0.421\pm0.01$ & $0.393\pm0.05$\tablenotemark{$^{b}$} & 0.53 \\
\cutinhead{planet e}
NIRC2-$L^{\prime}$ & $0.350\pm0.07$ & $0.395\pm0.055$\tablenotemark{$^{a}$} & -0.48 
\enddata
\tablenotetext{\dagger}{The uncertainty in the difference is taken to be the quadrature sum of the uncertainties on the individual measurements.}
\tablenotetext{a}{\citet{Currie2014}}
\tablenotetext{b}{\citet{Skemer2014}}
\end{deluxetable}

\section{Atmospheric model fitting}\label{sec:modelfitting}
\subsection{Compiling low-resolution data}
In combination with these ALES spectra, low resolution integral field
spectroscopy has measured the z, J, H, K, and L band emission from HR~8799~c,
d, and e. We fit model atmosphere spectra to these data, and, when available,
we include photometric measurements between and beyond the bands covered by
spectroscopy. For all the planets, measurements with a signal to noise less
than unity were clipped, and covariance matrices for IFS data were generated
following the approach outlined by \citet{Greco2016}. We assumed that the
significant binning required to produce the ALES synthetic photometry point
decoupled that point from the rest of the ALES spectra.

Below we briefly summarize the data compilations for each
planet.

\subsubsection{Planet c data} 

We combined our ALES measurements with the Project-1640 zJ band measurements
\citep[we used the version extracted with a PCA-based image post processing
algorithm][]{Oppenheimer2013} and included the GPI H and K band measurements
\citep{Greenbaum2018}. For the GPI measurements, we clipped data in the
overlapping region of the K1 and K2 filters (removing the last three points in
the K1 spectrum and the first eight points of K2).  The LBTI/LMIRCam LNB1,
LNB2, and LNB3 measurements from \citet{Skemer2014} were used in place of the
binned ALES measurements between 2.99~$\mu\mathrm{m}$ and 3.36~$\mu\mathrm{m}$
because they provide finer wavelength sampling and higher precision. The
Keck/NIRC2 $M$-band measurement from \citet{Galicher2011} was also included.

\subsubsection{Planet d data} 

For planet d, we combined our ALES measurements with the SPHERE IFS YH band
measurements \citep{Zurlo2016} and the GPI H and K-band measurements
\citep{Greenbaum2018}. We clipped the SPHERE data at the red end in order to
not overlap with the GPI H band measurements and clipped the GPI K1 and K2
spectra in the overlapping region, removing the last three points of K1 and the
first eight points of K2.  The LBTI/LMIRCam LNB1, LNB2, and LNB3 measurements
from \citet{Skemer2014} were used in place of the binned ALES measurements
between 2.99~$\mu\mathrm{m}$ and 3.36~$\mu\mathrm{m}$ because they provide
finer wavelength sampling and higher precision.The Keck/NIRC2 $M$-band
measurement from \citet{Galicher2011} were included.

\subsubsection{Planet e data} 

For planet e, we combined our ALES measurements with the SPHERE IFS YH band
measurements \citep{Zurlo2016} and the GPI H and K-band measurements
\citep{Greenbaum2018}. We clipped the SPHERE data at the red end in order to
not overlap with the GPI H band measurements, and clipped the GPI K1 and K2
spectra in the overlapping region, removing the last three points in of K1 and
the first eight points of K2. The 2.99~$\mu\mathrm{m}$ to 3.17~$\mu\mathrm{m}$
ALES synthetic photometry point is included, as this bin appears consistent
with previous observations for planets c and d. The
3.17~$\mu\mathrm{m}$ to 3.36~$\mu\mathrm{m}$ ALES synthetic photometry point is
  not used because this measurement appears to be affected by poor transmission
through the dgvAPP360.

\subsection{Fitting Approach} 

We fit synthetic spectra from three distinct families of models to the
measurements of each planet. The models were: 1) blackbodies; 2) DRIFT-Phoenix
models \citep{Witte2011}, which use a microphysics-based cloud prescription and
provide subsolar, solar, and supersolar metallicities; and 3) solar metallicity
Phoenix-based models with a parameterized cloud \citep[][]{Barman2015,
Brock2021}. {The $P_{\mathrm{c.t.}}$ parameter of the
Barman/Brock models is the pressure below which cloud particle density declines
exponentially. The median grain size and the eddy diffusion coefficient used
for the Barman/Brock models are $1~\mu\mathrm{m}$ and
$10^8~\mathrm{cm}^2\mathrm{s}^{-1}$, respectively. The parameter ranges and
step sizes for each grid are summarized in Table \ref{table_libraries}.}

The models were interpolated to provide finer sampling of their parameters
using multi-dimensional linear interpolation after rescaling input parameters
to the unit cube. For the synthetic atmosphere models we created 10~K steps in
effective temperature and steps of 0.1 dex in surface gravity. For the DRIFT
models, we created 0.1 dex steps in metallicity.  For the Barman/Brock models,
we created 0.3 dex steps in the pressure below which cloud particle density
decays exponentially. Blackbody models were precomputed with 2~K steps.

\begin{deluxetable*}{lcccc}[h]
\label{table_libraries}
\tablecaption{{Description of Model Libraries Used}}
\tablehead{\colhead{Parameter} &
\colhead{Barman/Brock} &
\colhead{DRIFT} &
\colhead{Blackbody}&
\colhead{Comments}
}
\startdata
$T_{\mathrm{eff}}$ range                   & 800--1500 K   & 1000--1500 K   & 800--1500 K & 100 K gridpoints interpolated to 10 K steps\\
$\log(\frac{g}{\mathrm{cm s}^{-2}})$ range & 3.5--5.0      & 3.5--5.0       & \nodata     & 0.5 dex gridpoints interpolated to steps of 0.1\\
$\log(\frac{z}{z_{\odot}})$ range          & \nodata       & -0.3--0.3      & \nodata     & 0.5 dex gridpoints interpolated to steps of 0.1\\
$P_{\mathrm{c.t.}}$\tablenotemark{*}       & 0.5, 1, 2, 4  & \nodata        & \nodata     & bars, the 2 bar model is interpolated\\
\enddata
\tablenotetext{*}{a log-uniform prior was used for $P_{\mathrm{c.t.}}$. Uniorm priors were used for all other parameters}
\end{deluxetable*}

\begin{deluxetable}{rccc}
\tabletypesize{\footnotesize}
\tablecolumns{4}
\tablewidth{0pt}
\tablecaption{Model Fits to HR~8799~c, d, e\label{modelsSummary}}
\tablehead{
    \colhead{Parameter}& 
    \colhead{planet c}&
    \colhead{planet d}&
    \colhead{planet e}
    }
\startdata
\cutinhead{Barman/Brock Phoenix Models}
$T_{\mathrm{eff}}$ [K]                                  & 1240 & 1140 & 1140 \\
$\log(\frac{g}{\mathrm{cm} \cdot \mathrm{s}^{2}})$      & 3.6  & 3.6  & 3.8  \\
$P_{\mathrm{ct}}$ [bar]                                 & 2    & 1    & 0.5   \\
$R$ [$R_\mathrm{Jup}$]                                  & 0.91 & 1.27 & 1.19 \\
$\log(\frac{L_{\mathrm{bol}}}{L_{\odot}})$              & -4.71& -4.62& -4.61 \\
$\chi^2$                                                & 234  & 1357 & 1379 \\
degrees of freedom                                      & 163  & 169  & 149 \\
\cutinhead{DRIFT Phoenix Models}
$T_{\mathrm{eff}}$ [K]                                  & 1500  & 1430 & 1480  \\
$\log(\frac{g}{\mathrm{cm} \cdot \mathrm{s}^{2}})$      & 3.5   & 3.5  & 3.9     \\
$\log(\frac{z}{z_{\odot}})$                             & 0.3   & -0.3 & -0.3  \\
$R$ [$R_\mathrm{Jup}$]                                  & 0.65  & 0.75 & 0.66  \\
$\log(\frac{L_{\mathrm{bol}}}{L_{\odot}})$              & -4.67 & -4.64& -4.67 \\
$\chi^2$                                                & 485  & 1933 & 1841 \\
degrees of freedom                                      & 163  & 169  & 149 \\
\cutinhead{Blackbody Models}
$T_{\mathrm{eff}}$ [K]                                  & 1424 & 1516 & 1620  \\
$R$ [$R_\mathrm{Jup}$]                                  & 0.75 & 0.62 & 0.5   \\
$\log(\frac{L_{\mathrm{bol}}}{L_{\odot}})$              & -4.65& -4.69& -4.70 \\
$\chi^2$                                                & 664  & 4323 & 3745 \\
degrees of freedom                                      & 165  & 171  & 151 \\
\enddata 
\end{deluxetable}

Prior to fitting, model spectra were preprocessed to match the characteristics
of each instrument.  This included smoothing to R=33 for fits to P1640 and
SPHERE spectra. For the GPI spectra we used the method presented in
\citet{Stone2016} to smooth the models with a linearly increasing spectral
resolution going from R=45 to R=80 from the beginning of H-band to the end of
K2. ALES data were fit with models smoothed to R=20. After smoothing, all model
spectra were sampled at the wavelengths provided by each instrument. We also
preprocessed photometry for the LMIRCam/LNB1, LNB2, LNB3, and the NIRC2
$M$-band points used. 

\begin{figure*}
\includegraphics[width=\linewidth]{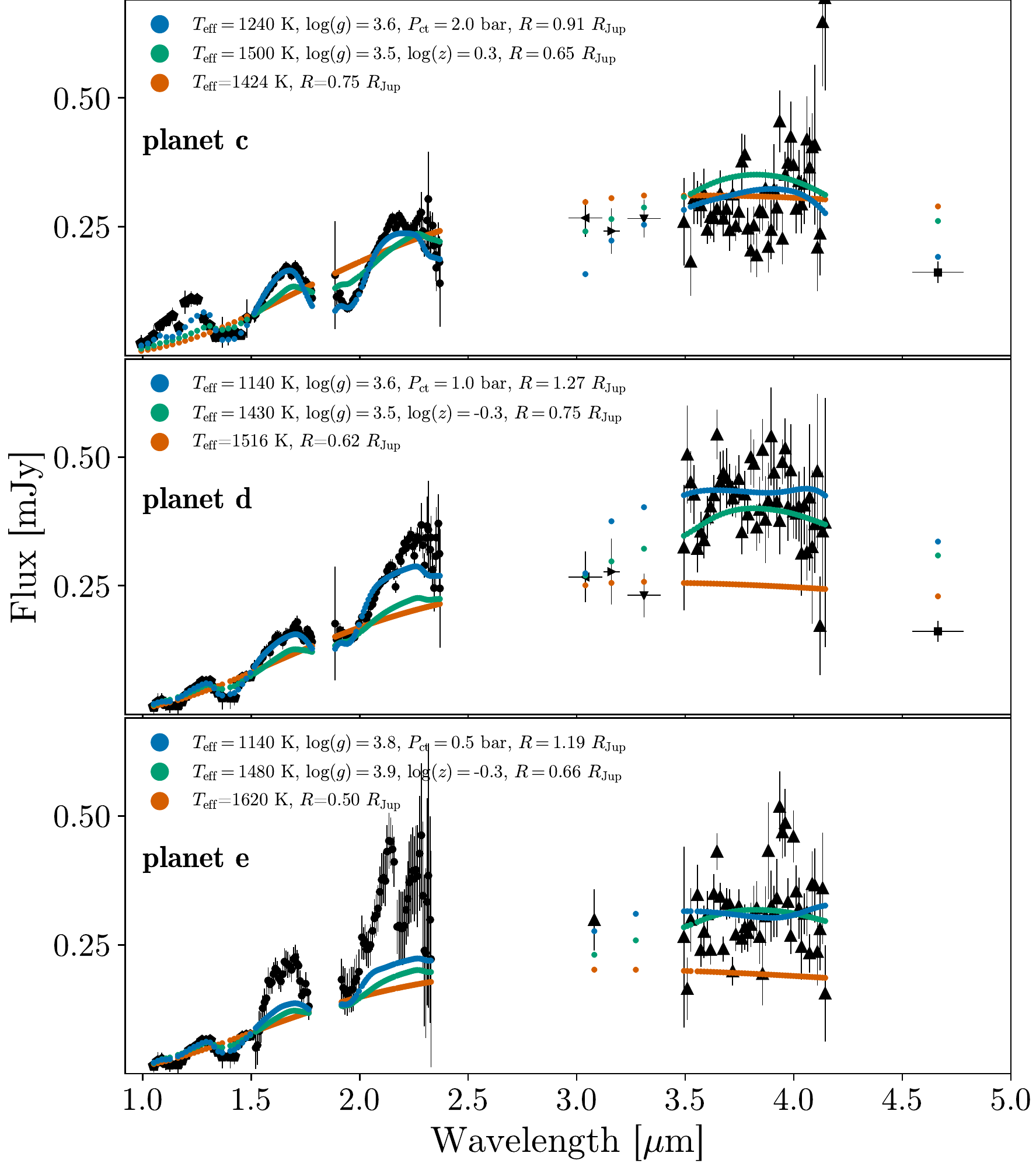}
\caption{The measured portions of the spectral energy distributions for planets
HR~8799~c, d, and e (black markers, data source-eps-converted-to.pdfs described in the text).
Colored markers indicate best fit model spectra from three model families. Blue
points are from the Barman/Brock Phoenix models. Teal points show DRIFT-Phoenix
models, and orange points are blackbodies. \label{HR8799columnFig}} 
\end{figure*}

Data were fit using a Gaussian likelihood function, treating each
band, $i$, individually
\begin{multline}
\mathcal{L}_{i}(\theta,R,\varpi)) \propto \\  
\mathrm{exp}\left(-\frac{1}{2}((R\varpi)^{2}x(\theta)-
\mu_{i})^{T}\Sigma_{i}^{-1}((R\varpi)^{2}x(\theta) - \mu_{i})\right),
\end{multline}
where $\theta$ represents a vector of model parameters for the given model
family, $R$ is the object radius, $\varpi$ is the system parallax, and
$\mu_{i}$ and $\Sigma_{i}$ are the measured data and covariance matrix for the
given spectrum or photometric measurement. We performed our fit using
a grid-based approach that facilitated the construction of a global likelihood
function through multiplication grid-cell by grid-cell
\begin{equation}
\mathcal{L}(\theta,R,\varpi) = \prod_{i}\mathcal{L}_{i}(\theta,R,\varpi)).
\end{equation}

We multiplied our likelihood grids by priors for each fitted parameter. For
each model family we used a log-uniform prior on $R$, extending from 0.5 to
2~$R_{\mathrm{Jup}}$, and a Gaussian prior on $\varpi$ using the $Gaia$
measurement and uncertainty {($24.46\pm0.045$ mas)}
as the mean and standard deviation, respectively \cite{GaiaCollaboration2018}.
For $T_{\mathrm{eff}}$, $\log(g)$, and $\log(z)$ a uniform prior was used. The
prior for $P_{\mathrm{c.t.}}$ was log-uniform.

\subsection{Model Fitting Results}\label{sec:Results}

Table \ref{modelsSummary} lists and Figure \ref{HR8799columnFig} displays our
fitting results. The Barman/Brock set of models, with greater cloud
flexibility, can provide reasonably close fits to the observations of planets
c and d. Neither the DRIFT models nor the Barman/Brock models fit planet
e particularly well, the H and K-bands being especially hard. As expected, the
black-body models provide a poor fit to the spectrum of each planet. 
For each planet, best fit models align most closely with the data having
smallest uncertainty, consistent with expectations. For HR~8799~c, since the
GPI data has the smallest uncertainty (and densest sampling), the optimal
models prefer to fit the H and K bands even if it costs a poorer fit through
the z, J, and L bands. For HR~8799~d, and e the SPHERE data has the smallest
uncertainty, so optimal models prefer to fit the z and J bands even if it costs
a poorer fit through the H, K, and L-bands.

Systematic differences between the models dominate our parameter uncertainty.
While within a model family, allowed parameter ranges
{(that is, the $\Delta\chi^2$=1 surface)} typically
span only one grid-cell, between the different models temperatures for planet
c span $\sim1200$ to $1500~\mathrm{K}$, temperatures for planet d span
$\sim1100$ to $1400~K$, and temperatures for planet e span $\sim1100$ to
$1600~\mathrm{K}$.  Surface gravity has less variance between the models,
constrained at the 0.1 dex level.  Best fit planet radii span 0.65 to
0.91~$R_{\mathrm{Jup}}$ for planet c, 0.62 to 1.27~$R_{\mathrm{Jup}}$ for
planet d, and 0.5 to 1.19~$R_{\mathrm{Jup}}$ for planet e.

Table \ref{modelsSummary} reports the inferred bolometric luminosity of each
planet. To derive the planet luminosity, a hybrid approach was employed
utilizing observed flux measurements wherever possible and integrating under
the best-fit model atmosphere at wavelengths between and beyond the measured
bands. {The uncertainty in the luminosity estimate is dominated by the
choice of atmospheric model family, yet the resulting values span only 0.09
dex, a very small uncertainty compared to the predictions of evolutionary
models ---resulting in a mass error of $\lesssim1~M_{\mathrm{Jup}}$ for a given
age, or an age error of less than 10~Myr for a given mass for the \citep[][See
Figure \ref{evoPlots}]{Baraffe2003}. The result suggests that luminosity is not
particularly sensitive to the choice of well scaled model, especially in this
case where we have broad wavelength coverage near the peak of each planet's
spectral energy distribution.}

\section{Discussion} \label{sec:discussion} 
Hot-start evolutionary models predict a very narrow range of effective
temperatures, surface gravity, and planetary radius constrained by the
fundamental parameters of the HR~8799 planets. Constrained parameters include
system age \citep[$33^{+7}_{-13}$ Myr or $90^{381}_{-50}$ Myr,][]{Baines2012},
planet mass \citep[planets c and d $\lesssim10~M_{\mathrm{Jup}}$, planet
e $9.6^{+1.9}_{-1.8}~M_{\mathrm{Jup}}$,][]{Fabrycky2010, Brandt2021}, and
bolometric luminosity (Table \ref{modelsSummary}). 
\begin{figure*}
\includegraphics[width=\linewidth]{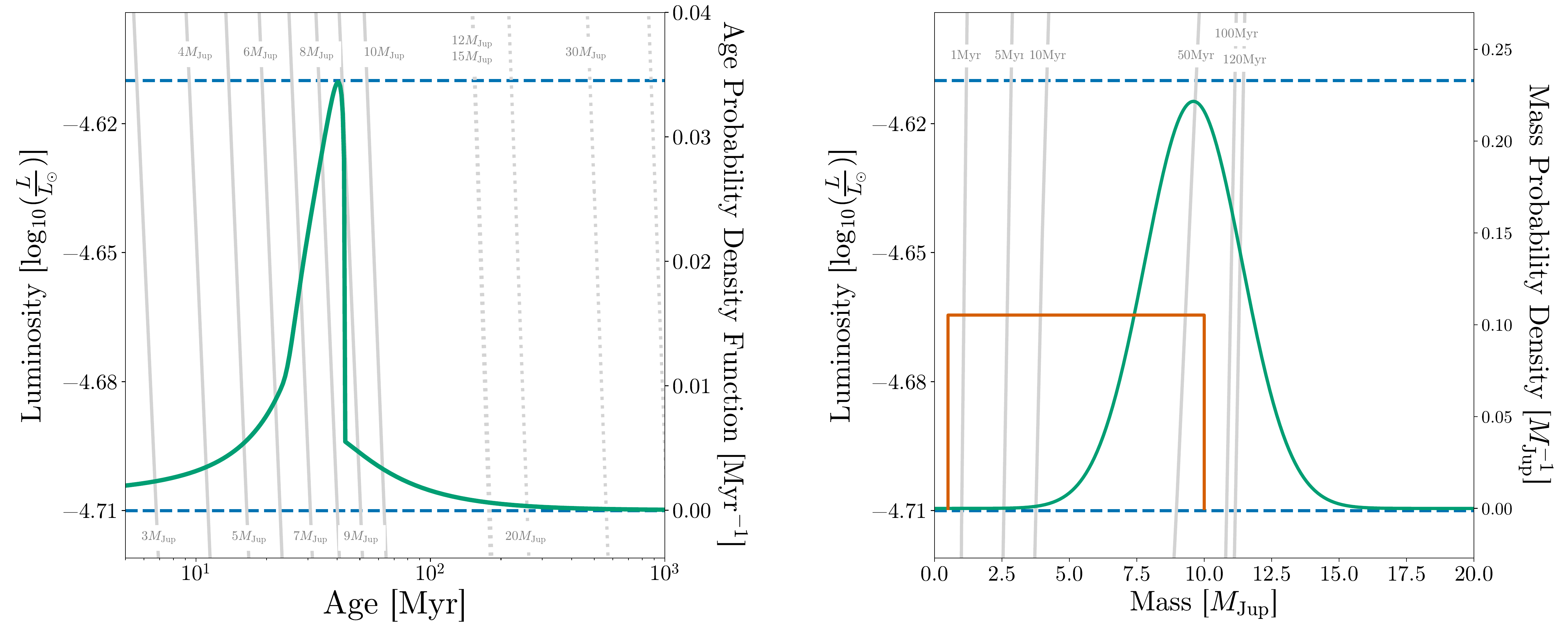}
\caption{{Left: Luminosity versus age. The gray
curves represent the predictions of the \citet{Baraffe2003} evolutionary
models. Blue horizontal lines indicate the luminosity constraints for the
HR~8799 planets. The green curve corresponds to the right axis and shows the
system age probability distribution function giving equal weight to the
contracting and expanding scenarios from \citet{Baines2012}. Right: Luminosity
versus mass. Gray curves are from the evolution model. Blue horizontal lines
indicate the luminosity constraints. The orange and green curves correspond to
the right axis and indicate probability distribution functions derived from
stability arguments (orange) and from the astrometric influence of planet
e (green).}\label{evoPlots}} \end{figure*}

To illustrate this we used the `evolve' module of the SpeX Prism Library
Analysis Toolkit \citep[SPLAT,][]{Burgasser2017} to construct a distribution of
evolution model predictions for effective temperature, surface gravity, and
radius. We used a Monte Carlo approach to build distributions for four
different models \citep{Burrows2001, Saumon2008, Baraffe2003}. 1.2 million
age-mass points were input into the evolutionary models and the output
discarded if the returned luminosity was outside the measured range. For the
system age, we modeled each of the ranges indicated by \citet{Baines2012} using
a generalized extreme value distribution \citep{Possolo2019}, giving equal
weight to the younger and the older ranges.  Planet masses were sampled from
a uniform distribution spanning 0.5 to 10~$M_{\mathrm{Jup}}$,
{consistent with dynamical constraints
\citep[e.g.,][]{Fabrycky2010}}. Since the allowed luminosity ranges for each
planet are similar, we used a single range for all planets, spanning
$\log_{10}(\frac{L_{\mathrm{bol}}}{L_{\odot}})$=-4.71 to -4.61.
{The intersection of these constraints on the
\citet{Baraffe2003} evolutionary models are plotted in Figure \ref{evoPlots} as
an example.}

\begin{figure}
\includegraphics[width=\linewidth]{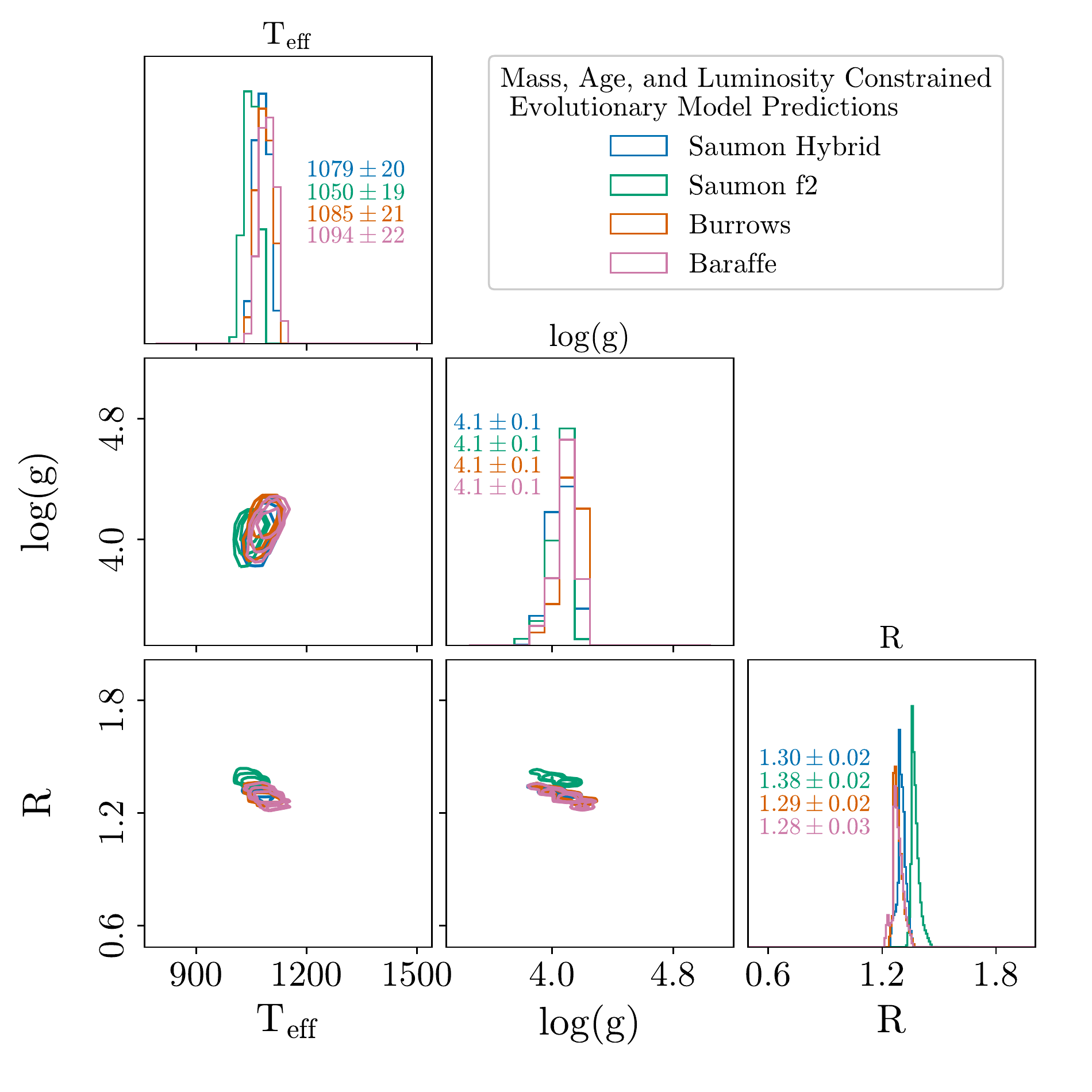}
\caption{Temperature, surface gravity, and radius predictions from four
hot-start evolutionary models, given the age, mass, and luminosity constraints
for HR~8799~c, d, and e. Predictions are narrowly peaked and are relatively
model independent.  \label{evopredictionsFig}} \end{figure}

The results of our Monte Carlo sampling are displayed in Figure
\ref{evopredictionsFig}. We repeated the sampling exercise using a Gaussian
distributed mass constraint approximating the results of
\citet[][{$M=9.6\pm1.8~M_{\mathrm{Jup}}$}]{Brandt2021}
and no significant change to the resulting distributions resulted. 

The predictions of gas-giant evolution models are sensitive to the initial
entropy assumed during early times.  For 10~$M_{\mathrm{Jup}}$ objects
hot-start and cold-start evolutionary models do not converge for $\sim1$~Gyr
\citep[less massive objects converge faster][]{Marley2007}. Each of the four
models we use assume hot-start evolution.  Hot start models 
are consistent with initial entropy constraints for the planets
\citet{Marleau2014}, however "warm"-start models are also allowed. 

Assuming hot-start evolution, Figure \ref{evopredictionsFig} suggests
$T_{\mathrm{eff}}\approx1075$~K, $\log(g)\approx4.1$, and
$R\approx1.29~R_{\mathrm{Jup}}$. Comparing to Table \ref{modelsSummary} we see
that the best-fit Barman/Brock Phoenix models provide reasonable parameters for
HR~8799~d and HR~8799~e, with $T_{\mathrm{eff}}$ within 100~K, $\log(g)$ within
0.5 dex, and plausible planet radii. The best-fit for HR~8799~c has
  more tension with the evolutionary models, suggesting a radius of
0.91~$R_{\mathrm{Jup}}$. The best-fit DRIFT models have temperatures much
higher (and radii much smaller) than predicted by evolution models.

\begin{figure*}
\includegraphics[width=\linewidth]{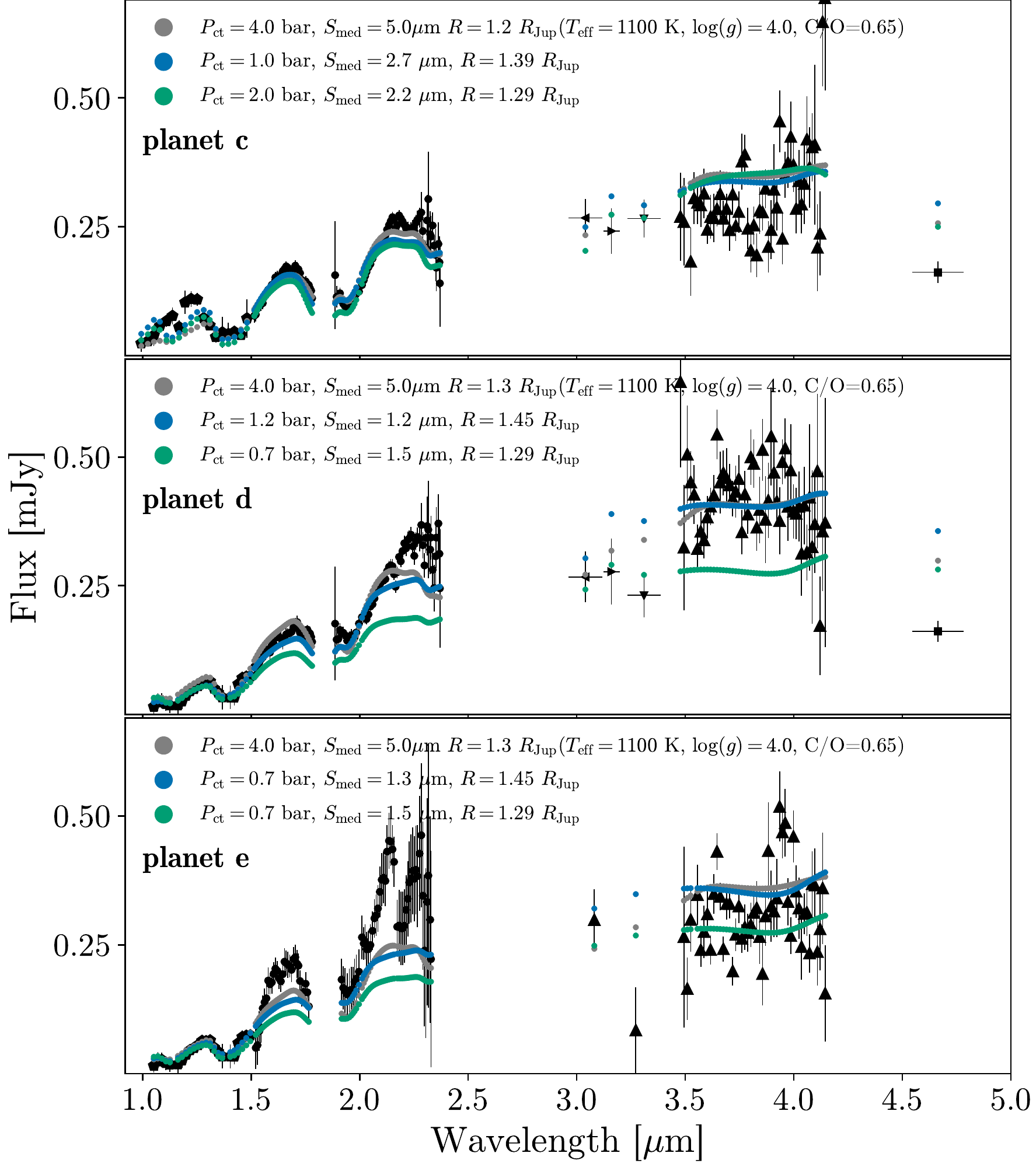}
\caption{The measured portions of the spectral energy distributions for planets
HR~8799~c, d, and e (black markers, data sources described in the text).
Barman/Brock Phoenix models with temperature and surface gravity fixed at the
values predicted by hot-start evolutionary models ($T_{\mathrm{eff}}=1075$~K,
$\log(g)=4.1$) are fit to the data by varying two cloud parameters. Blue points
are for a fit letting the radius float between 1.25 and 1.45
$R_{\mathrm{Jup}}$. Teal points show a fit where radius is fixed to
$1.29~R_{\mathrm{Jup}}$. Thick clouds with small grains are preferred for each
planet, suggesting such conditions are necessary to bring atmospheric models
and evolutionary models into agreement. We also show the
$T_{\mathrm{eff}}=1100$~K, $\log(g)=4$, Phoenix model presented by
\citet{Konopacky2013} and \citet{Greenbaum2018} (gray markers). This model is
also consistent with the evolutionary model predictions. It was drawn from the
same Barman/Brock family of models presented here but includes a supersolar C/O
ratio.  \label{EvocolumnFig}} \end{figure*}

Given the narrowly peaked distributions in Figure \ref{evopredictionsFig}, we
re-ran our fitter, fixing $T_{\mathrm{eff}}=1075$~K and $\log(g)$=4.1. We fit
twice, using different constraints on planet radius each time. First, we
restricted $R$ to be within 1.25 and 1.45~$R_{\mathrm{Jup}}$. Second we fixed
$R$ at 1.29~$R_{\mathrm{Jup}}$. We also used more flexible models, now exposing
both cloud top pressure and median grain size as tunable parameters. Our
question was: Assuming evolutionary models are correct, what does that imply
about the physical state of the atmosphere? The results are shown in Figure
\ref{EvocolumnFig}.

By adjusting both the cloud top pressure and the median grain size, we are able
to find reasonably good fits to the data at the expected temperature and
surface gravity. In all cases, small median grain-size, $<3~\mu\mathrm{m}$, are
preferred. HR~8799~e is fit with the most extended cloud, but this planet is
also the least well matched by the models, the H and K-band spectra are
systematically under predicted. Given that HR~8799~e is the closest-in and
hardest to observe planet, it is hard to distinguish persistent systematic
issues with the spectroscopy from deficiencies of the model atmospheres and
probably both are at play. For HR~8799~d, as seen in the middle panel of Figure
\ref{EvocolumnFig}, the fixed $R=1.29~R_{\mathrm{Jup}}$ fit is a poor match to
the data, especially compared to the more flexible constrained fit and the
unconstrained fit. 

{Additional atmospheric model parameters are necessary to provide improved fits.
In \citet{Konopacky2013} a Phoenix model from the same family as the
Barman/Brock models used here is fit to higher resolution K-band spectroscopy
of HR~8799~c. The atmospheric model, with $T_{\mathrm{eff}}=1100$~K and
$\log(g)=4.0$, is consistent with the predictions of hot-start evolution
models, and matches low-resolution spectroscopy and the broadband SED reasonably
well \citep{Greenbaum2018}. In addition to cloud top pressure and median grain
size, the \citet{Konopacky2013} model also adjusts the C/O ratio in the
planetary atmosphere. Figure \ref{EvocolumnFig} compares our results with the
Phoenix model of \citet{Konopacky2013}.}

{Our analysis does not utilize models with a cloud coverage/patchiness
parameter, yet some previous studies suggest that this may be important for the
HR~8799 planets \citep[e.g.,][]{Currie2014, Skemer2014}. For HR~8799~c, our
fits in Figures \ref{HR8799columnFig} and \ref{EvocolumnFig} show that matching
the zJ band emission while simultaneously fitting longer wavelengths is
challenging. One way to enhance emission from the z through J bands is with
cloud patches \citep{Marley2010}.}

\section{Conclusion}\label{sec:conclusion} 

New coronagraphic L-band integral field spectroscopy of the HR~8799 system are
presented using the LMIRCam/ALES instrument and a double-grating vector
apodizing phase plate.  These are the first L-band spectroscopic measurements
of HR~8799~d and e, and the first broadband spectroscopy of HR~8799~c at these
wavelengths. Our measurements are {generally} consistent with earlier
photometric probes covering portions of this band, {although there is some
tension with \citet{Skemer2014} for HR~8799~c.}

Atmospheric model fits incorporating a parameterized treatment of clouds can
provide a reasonable fit to the 1-4.6~$\mu\mathrm{m}$ spectral energy
distribution of the planets when the median grain size is small.  HR~8799~d is
particularly well fit with atmospheric models that agree with the predictions
of hot-start evolutionary models.

An approach of sampling evolution models is developed to create distributions
of predicted temperature, gravity, and radius given constraints on several
fundamental parameters of the HR~8799 system. Re-running of fits fixing these
parameters enables the use of a more flexible synthetic atmosphere model.

While evolution models depend on initial conditions for ages $\lesssim1$~Gy,
our approach can be applied to older systems where ambiguity about the
formation path is less of an issue. Powerful thermal-IR instruments, such as
LMIRCam/ALES, are capable of direct imaging observations of older planets,
which maintain their thermal-IR flux even as the near-IR fades dramatically.
Future facilities like JWST and the next generation of 30-m class telescopes
with mid-IR instruments such as METIS \citep{Brandl2021} and PSI-red
\citep{Skemer2018b} will extend this capability to fainter/older targets.  When
paired with the power of Gaia to find and constrain the masses
of wide orbit planets, in future studies we will be well equipped to make use
of evolutionary models to provide quantitative priors for their atmospheric
model fitting.

\appendix

\section{Characterization of the dgvAPP360.} \label{sec:vAPP_characterization}

The double grating vector apodizing phase plate 360 (dgvAPP360) was installed
in LMIRCam early September 2018. The design and first-light results are
presented in \cite{Doelman2020}.  Observations of the PDS 201 system using this
dgvAPP360 show that the vAPP has an improved sensitivity of a factor two
compared to non-coronagraphic imaging in the regions closest to the star
(450-800 mas) \citep{Wagner2020}. Here we provide further characterization of
the dgvAPP360 performance, focusing on the throughput as a function of
wavelength.

\begin{figure}
    \centering
    \includegraphics[width=0.7\linewidth]{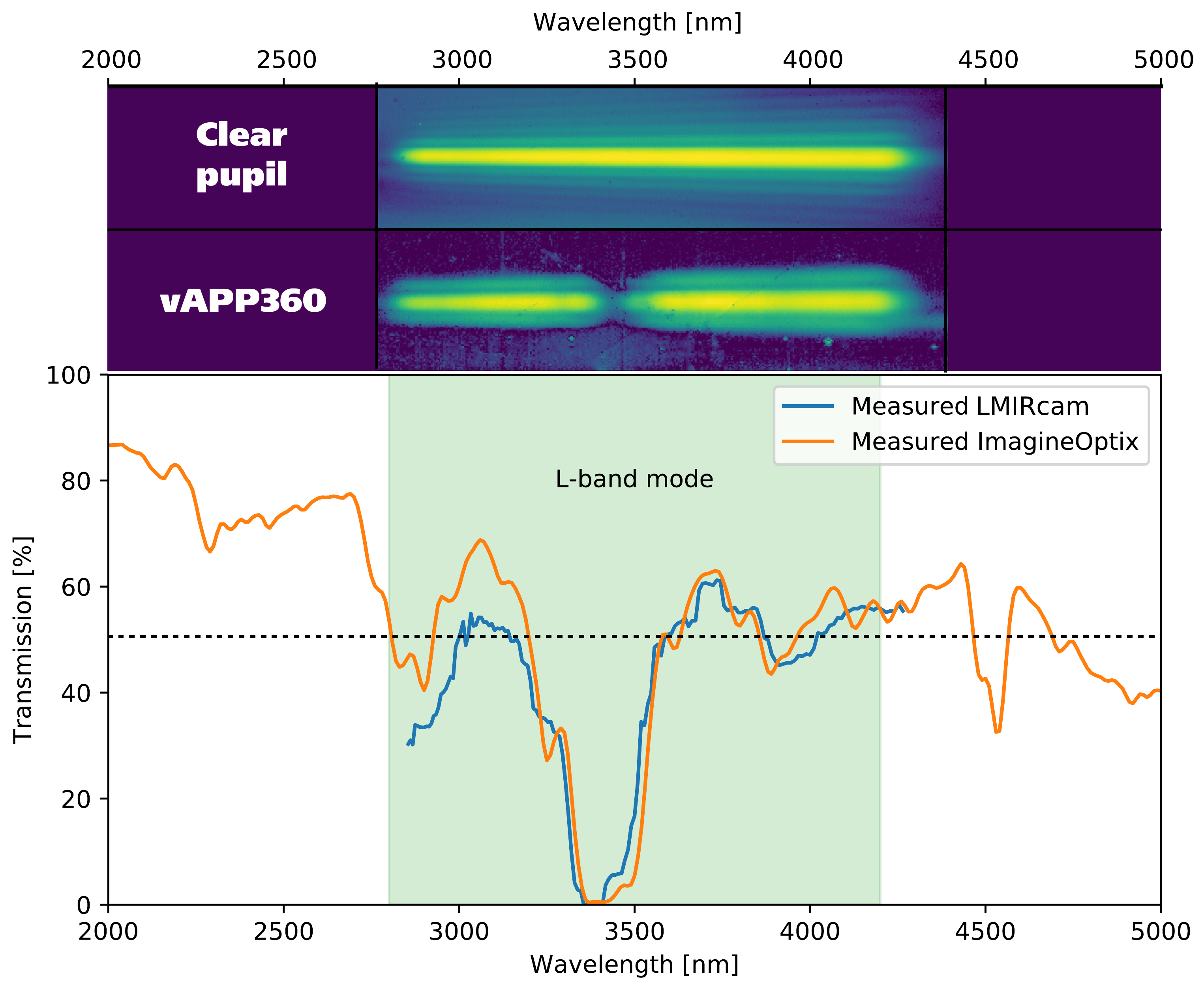}
    \caption{Transmission measurements of the vAPP in the 2.8-4.2 $\mu$m range
    using a grism to disperse the vAPP and non-coronagraphic PSF.}
    \label{fig:throughput}
\end{figure}

The double-grating vAPP has two separate liquid-crystal layers, an additional
glue layer and an extra substrate compared to a standard gvAPP
\citep{Doelman2017,Doelman2020}.  These additional layers lead to extra
absorption, specifically in the $3.24$-$3.5$ $\mu$m range, where both the
liquid-crystal molecules and the glue molecules have an infra-red absorption
feature due to carbon-carbon bonds \citep{Otten2017}. 
Because these features are in the spectral range of ALES, we conduct an
experiment to measure their impact.  We disperse the coronagraphic (vAPP) PSF
and non-coronagraphic (clear pupil) PSF using a grism in tandem with the
$L$-Spec filter, using only the left aperture (SX) of the internal pupil mask.
The dispersed PSFs are shown in Fig. \ref{fig:throughput}.  

In addition, we obtain narrow band images at $2.9$ $\mu$m, $3.3$ $\mu$m, $3.5$
$\mu$m, and $3.9$ $\mu$m that are used for wavelength calibration.  Because the
pupil selection mask is in the same filter wheel as the narrow band filters,
the narrow band images are the coherent sum of both pupils, creating Fizeau
fringes.  We remove this effect for our wavelength calibration by summing the
flux in 100 pixels in the fringe direction.  We fit a Gaussian to the
one-dimensional sum of each wavelength to retrieve accurate centroids.  We then
use the centroids to calculate the wavelength solution of the dispersion.

Calculating the transmission of the vAPP is complicated by the difference in
intrinsic Strehl between the two PSFs.  The vAPP has an intrinsic Strehl of
46\% and the PSF core is slightly broadened due to the apodization by the vAPP
in the pupil plane.  We retrieve the throughput by forward modelling both PSFs
for the full bandwidth.  For continuous wavelength coverage, we generate
a wavelength-scaled PSF every 2 pixels in the image, corresponding to an
average spectral resolution of 10 nm.  The model PSF is the incoherent sum of
all individual PSFs with an individual scale factor.  We minimize the
difference between the simulated and measured PSF by changing these scale
factors, thereby retrieving the true input spectrum.  The final transmission is
the optimized vAPP spectrum divided by the non-coronagraphic spectrum, and is
shown in Fig. \ref{fig:throughput}.  We compare the results with the
transmission that is measured by the manufacturer.  The curves are in good
agreement except for the transmission between 2.9 and 3.2 micron.  It is
unclear what causes the difference of more than 10\% in this spectral bin,
compared to the LMIRCam average transmission of 47\%.  A significant fraction
of the spectral band is unavailable when using the vAPP because of the
absorption feature.  Specifically, the absorption is centered on a spectral
feature of CH$_4$.  While this inhibits the detection of methane in the
atmospheres of L-type gas giants, cooler gas giants with more CH$_4$ will have
a measurable spectral slope between 3.5-4.1 $\mu$m, where the average
transmission of the vAPP is 51\%.
\newpage
\section{Tabulated Planet Spectra}
\label{App:table_spectra}
\startlongtable
\begin{deluxetable}{ccccccc}
\tablecaption{ALES spectra of HR~8799~c, d, and e}
\tablehead{\colhead{Wavelength} &
           \colhead{Planet c flux} &
           \colhead{Planet c flux uncertainty} &
           \colhead{Planet d flux} &
           \colhead{Planet d flux uncertainty} &
           \colhead{Planet e flux} &
           \colhead{Planet e flux uncertainty} \\
           \colhead{(micons)} &
           \colhead{(mJy)} &
           \colhead{(mJy)} &
           \colhead{(mJy)} &
           \colhead{(mJy)} &
           \colhead{(mJy)} &
           \colhead{(mJy)} }

\startdata
3.081 & 0.2078 & 0.0536 & 0.3101 & 0.0713 & 0.2984 & 0.0597 \\
3.272 & 0.1115 & 0.0614 & 0.6491 & 0.1169 & 0.0842 & 0.0831 \\
3.493 & 0.2594 & 0.0848 & 0.3242 & 0.1215 & 0.2654 & 0.1754\\
3.509 & 0.0861 & 0.0892 & 0.5048 & 0.0953 & 0.1647 & 0.0605 \\
3.525 & 0.1822 & 0.0659 & 0.4510 & 0.0598 & 0.2986 & 0.0615 \\
3.540 & 0.3052 & 0.0503 & 0.4271 & 0.0571 & 0.0333 & 0.0557 \\
3.556 & 0.2967 & 0.0420 & 0.3218 & 0.0464 & 0.3469 & 0.0581 \\
3.571 & 0.2916 & 0.0422 & 0.3553 & 0.0383 & 0.2400 & 0.0325 \\
3.586 & 0.3143 & 0.0479 & 0.3385 & 0.0353 & 0.2753 & 0.0513 \\
3.601 & 0.2443 & 0.0279 & 0.3830 & 0.0415 & 0.3096 & 0.0450 \\
3.616 & 0.2690 & 0.0425 & 0.4038 & 0.0364 & 0.2403 & 0.0284 \\
3.631 & 0.2668 & 0.0467 & 0.4258 & 0.0389 & 0.3487 & 0.0381 \\
3.646 & 0.2842 & 0.0338 & 0.5437 & 0.0484 & 0.4311 & 0.0355 \\
3.660 & 0.3135 & 0.0456 & 0.4506 & 0.0598 & 0.3433 & 0.0265 \\
3.675 & 0.2659 & 0.0346 & 0.4689 & 0.0432 & 0.2420 & 0.0241 \\
3.689 & 0.2846 & 0.0290 & 0.4590 & 0.0586 & 0.3296 & 0.0301 \\
3.704 & 0.2434 & 0.0406 & 0.4452 & 0.0412 & 0.3295 & 0.0282 \\
3.718 & 0.3126 & 0.0391 & 0.4194 & 0.0345 & 0.1988 & 0.0276 \\
3.732 & 0.2514 & 0.0290 & 0.4327 & 0.0508 & 0.2697 & 0.0317 \\
3.746 & 0.2788 & 0.0402 & 0.4573 & 0.0465 & 0.3248 & 0.0325 \\
3.760 & 0.3766 & 0.0545 & 0.3540 & 0.0426 & 0.2621 & 0.0272 \\
3.774 & 0.3901 & 0.0382 & 0.4274 & 0.0459 & 0.2842 & 0.0439 \\
3.788 & 0.2463 & 0.0402 & 0.3886 & 0.0379 & 0.2732 & 0.0369 \\
3.802 & 0.2038 & 0.0426 & 0.4995 & 0.0533 & 0.2888 & 0.0431 \\
3.815 & 0.2525 & 0.0369 & 0.4869 & 0.0466 & 0.3122 & 0.0526 \\
3.829 & 0.1946 & 0.0421 & 0.3630 & 0.0450 & 0.3213 & 0.0520 \\
3.842 & 0.2813 & 0.0454 & 0.3975 & 0.0645 & 0.2657 & 0.0525 \\
3.856 & 0.2780 & 0.0356 & 0.5138 & 0.0601 & 0.1936 & 0.0608 \\
3.869 & 0.3239 & 0.0379 & 0.3785 & 0.0845 & 0.3054 & 0.0575 \\
3.882 & 0.2110 & 0.0498 & 0.4156 & 0.0730 & 0.4325 & 0.0938 \\
3.895 & 0.2439 & 0.0571 & 0.5400 & 0.0953 & 0.3248 & 0.0689 \\
3.908 & 0.3220 & 0.0876 & 0.4696 & 0.0619 & 0.3148 & 0.0500 \\
3.921 & 0.2872 & 0.0450 & 0.4135 & 0.0429 & 0.3405 & 0.0760 \\
3.934 & 0.4545 & 0.0598 & 0.3760 & 0.0647 & 0.5182 & 0.0680 \\
3.947 & 0.2267 & 0.0495 & 0.4902 & 0.0434 & 0.4687 & 0.0509 \\
3.960 & 0.3507 & 0.0442 & 0.5168 & 0.0584 & 0.4867 & 0.0654 \\
3.973 & 0.3741 & 0.0485 & 0.4035 & 0.0659 & 0.3334 & 0.0456 \\
3.985 & 0.4246 & 0.0678 & 0.4740 & 0.0892 & 0.2675 & 0.0354 \\
3.998 & 0.3697 & 0.0523 & 0.3944 & 0.0522 & 0.4600 & 0.0503 \\
4.010 & 0.2847 & 0.0473 & 0.3899 & 0.0497 & 0.3536 & 0.0508 \\
4.023 & 0.3388 & 0.0606 & 0.4036 & 0.0679 & 0.3208 & 0.0597 \\
4.035 & 0.2933 & 0.0566 & 0.3124 & 0.0819 & 0.2460 & 0.0642 \\
4.048 & 0.3333 & 0.0785 & 0.4057 & 0.0755 & 0.3103 & 0.0398 \\
4.060 & 0.4194 & 0.0842 & 0.3149 & 0.0543 & 0.3127 & 0.0459 \\
4.072 & 0.3649 & 0.0785 & 0.4215 & 0.1421 & 0.2336 & 0.0372 \\
4.084 & 0.4044 & 0.0662 & 0.3252 & 0.1486 & 0.3682 & 0.0635 \\
4.096 & 0.4092 & 0.1552 & 0.3696 & 0.0943 & 0.3655 & 0.0716 \\
4.108 & 0.2094 & 0.0841 & 0.4724 & 0.1497 & 0.2363 & 0.0649 \\
4.120 & 0.2367 & 0.0817 & 0.1724 & 0.0959 & 0.2806 & 0.0799 \\
4.132 & 0.6473 & 0.1246 & 0.3555 & 0.1159 & 0.3597 & 0.1081\\
4.144 & 0.7018 & 0.1879 & 0.3727 & 0.2420 & 0.1557 & 0.0930 \\
\enddata

\end{deluxetable}

\begin{deluxetable}{cccccc}
\tablecaption{Spectral correlation parameters for ALES spectra of HR~8799 planets}
\tablehead{\colhead{Planet} &
           \colhead{$A_{\rho}$} &
           \colhead{$\sigma_{\rho}$} &
           \colhead{$A_{\lambda}$} &
           \colhead{$\sigma_{\lambda}$} &
           \colhead{$A_{\delta}$} \\
           \colhead{} &
           \colhead{} &
           \colhead{} &
           \colhead{} &
           \colhead{} &
           \colhead{} }

\startdata
c & 1.24E-2 & 1.61E02 & 3.95E-1 & 2.42E-3 & 5.94E-1 \\
d & 1.17E-2 & 1.02E03 & 4.09E-1 & 2.47E-3 & 5.81E-1 \\
e & 2.92E-2 & 1.36E04 & 1.24E-1 & 2.57E-2 & 8.47E-1 \\
\enddata
\tablecomments{Parameters as defined by \citet{Greco2016}}
\end{deluxetable}

\begin{acknowledgements}
We thank Rebecca Oppenheimer, Laurent Pueyo, Alex Greenbaum, Alice Zurlo, and
Dino Mesa for making their data available so that we could construct covariance
matrices for use in model fitting to IFS data.

This work benefited from the Exoplanet Summer Program in
the Other Worlds Laboratory (OWL) at the University of California, Santa Cruz,
a program funded by the Heising-Simons Foundation.  

This paper is based on work funded by NSF Grants 1608834, 1614320 and 1614492.
Work conducted by Laci Brock and Travis Barman was also supported by the
National Science Foundation under Award No. 1405504.  J.M.S. was partially
supported by NASA through Hubble Fellowship grant HST-HF2-51398.001-A awarded
by the Space Telescope Science Institute, which is operated by the Association
of Universities for Research in Astronomy, Inc., for NASA, under contract
NAS5-26555.  Z.W.B. is supported by the National Science Foundation Graduate
Research Fellowship under Grant No. 1842400. C.E.W acknowledges partial support
from NASA grant 80NSSC19K0868.

The LBT is an international collaboration among institutions in the United
States, Italy and Germany.  LBT Corporation partners ar: The University of
Arizona on behalf of the Arizona university system; Istituto Nazionale di
Astrofisica, Italy; LBT Beteiligungsgesellschaft, Germany, representing the
Max-Planck Society, the Astrophysical Institute Potsdam, and Heidelberg
University; The Ohio State University, and The Research Corporation, on behalf
of The University of Notre Dame, University of Minnesota, and University of
Virginia.  We thank all LBTI team members for their efforts that enabled this
work.    
\end{acknowledgements}

\facilities{LBT (LBTI/LMIRCam, LBTI/ALES)}

\software{Astropy \citep{astropy2013},
Matplotlib \citep{matplotlib},
Scipy \citep{2020SciPy-NMeth},
SPLAT \citep{Burgasser2017}),
}

\bibliography{HR8799.bib}

\begin{thebibliography}{}
\expandafter\ifx\csname natexlab\endcsname\relax\def\natexlab#1{#1}\fi

\bibitem[{{Astropy Collaboration} {et~al.}(2013){Astropy Collaboration},
  {Robitaille}, {Tollerud}, {Greenfield}, {Droettboom}, {Bray}, {Aldcroft},
  {Davis}, {Ginsburg}, {Price-Whelan}, {Kerzendorf}, {Conley}, {Crighton},
  {Barbary}, {Muna}, {Ferguson}, {Grollier}, {Parikh}, {Nair}, {Unther},
  {Deil}, {Woillez}, {Conseil}, {Kramer}, {Turner}, {Singer}, {Fox}, {Weaver},
  {Zabalza}, {Edwards}, {Azalee Bostroem}, {Burke}, {Casey}, {Crawford},
  {Dencheva}, {Ely}, {Jenness}, {Labrie}, {Lim}, {Pierfederici}, {Pontzen},
  {Ptak}, {Refsdal}, {Servillat}, \& {Streicher}}]{astropy2013}
{Astropy Collaboration}, {Robitaille}, T.~P., {Tollerud}, E.~J., {et~al.} 2013,
  \aap, 558, A33

\bibitem[{{Baines} {et~al.}(2012){Baines}, {White}, {Huber}, {Jones},
  {Boyajian}, {McAlister}, {ten Brummelaar}, {Turner}, {Sturmann}, {Sturmann},
  {Goldfinger}, {Farrington}, {Riedel}, {Ireland}, {von Braun}, \&
  {Ridgway}}]{Baines2012}
{Baines}, E.~K., {White}, R.~J., {Huber}, D., {et~al.} 2012, \apj, 761, 57

\bibitem[{{Baraffe} {et~al.}(2003){Baraffe}, {Chabrier}, {Barman}, {Allard}, \&
  {Hauschildt}}]{Baraffe2003}
{Baraffe}, I., {Chabrier}, G., {Barman}, T.~S., {Allard}, F., \& {Hauschildt},
  P.~H. 2003, \aap, 402, 701

\bibitem[{{Barman} {et~al.}(2015){Barman}, {Konopacky}, {Macintosh}, \&
  {Marois}}]{Barman2015}
{Barman}, T.~S., {Konopacky}, Q.~M., {Macintosh}, B., \& {Marois}, C. 2015,
  \apj, 804, 61

\bibitem[{{Barman} {et~al.}(2011){Barman}, {Macintosh}, {Konopacky}, \&
  {Marois}}]{Barman2011a}
{Barman}, T.~S., {Macintosh}, B., {Konopacky}, Q.~M., \& {Marois}, C. 2011,
  \apj, 733, 65

\bibitem[{{Bayo} {et~al.}(2008){Bayo}, {Rodrigo}, {Barrado Y Navascu{\'e}s},
  {Solano}, {Guti{\'e}rrez}, {Morales-Calder{\'o}n}, \& {Allard}}]{Bayo2008}
{Bayo}, A., {Rodrigo}, C., {Barrado Y Navascu{\'e}s}, D., {et~al.} 2008, \aap,
  492, 277

\bibitem[{{Bonnefoy} {et~al.}(2016){Bonnefoy}, {Zurlo}, {Baudino}, {Lucas},
  {Mesa}, {Maire}, {Vigan}, {Galicher}, {Homeier}, {Marocco}, {Gratton},
  {Chauvin}, {Allard}, {Desidera}, {Kasper}, {Moutou}, {Lagrange}, {Antichi},
  {Baruffolo}, {Baudrand}, {Beuzit}, {Boccaletti}, {Cantalloube}, {Carbillet},
  {Charton}, {Claudi}, {Costille}, {Dohlen}, {Dominik}, {Fantinel},
  {Feautrier}, {Feldt}, {Fusco}, {Gigan}, {Girard}, {Gluck}, {Gry}, {Henning},
  {Janson}, {Langlois}, {Madec}, {Magnard}, {Maurel}, {Mawet}, {Meyer},
  {Milli}, {Moeller-Nilsson}, {Mouillet}, {Pavlov}, {Perret}, {Pujet}, {Quanz},
  {Rochat}, {Rousset}, {Roux}, {Salasnich}, {Salter}, {Sauvage}, {Schmid},
  {Sevin}, {Soenke}, {Stadler}, {Turatto}, {Udry}, {Vakili}, {Wahhaj}, \&
  {Wildi}}]{Bonnefoy2016}
{Bonnefoy}, M., {Zurlo}, A., {Baudino}, J.~L., {et~al.} 2016, \aap, 587, A58

\bibitem[{{Brandl} {et~al.}(2021){Brandl}, {Bettonvil}, {van Boekel},
  {Glauser}, {Quanz}, {Absil}, {Amorim}, {Feldt}, {Glasse}, {G{\"u}del}, {Ho},
  {Labadie}, {Meyer}, {Pantin}, {van Winckel}, \& {METIS
  Consortium}}]{Brandl2021}
{Brandl}, B., {Bettonvil}, F., {van Boekel}, R., {et~al.} 2021, The Messenger,
  182, 22

\bibitem[{{Brandt} {et~al.}(2021){Brandt}, {Brandt}, {Dupuy}, {Michalik}, \&
  {Marleau}}]{Brandt2021}
{Brandt}, G.~M., {Brandt}, T.~D., {Dupuy}, T.~J., {Michalik}, D., \& {Marleau},
  G.-D. 2021, \apjl, 915, L16

\bibitem[{{Briesemeister} {et~al.}(2018){Briesemeister}, {Skemer}, {Stone},
  {Stelter}, {Hinz}, {Leisenring}, {Skrutskie}, {Woodward}, \&
  {Barman}}]{Briesemeister2018}
{Briesemeister}, Z., {Skemer}, A.~J., {Stone}, J.~M., {et~al.} 2018, in Society
  of Photo-Optical Instrumentation Engineers (SPIE) Conference Series, Vol.
  10702, Ground-based and Airborne Instrumentation for Astronomy VII, ed. C.~J.
  {Evans}, L.~{Simard}, \& H.~{Takami}, 107022Q

\bibitem[{{Brock} {et~al.}(2021){Brock}, {Barman}, {Konopacky}, \&
  {Stone}}]{Brock2021}
{Brock}, L., {Barman}, T., {Konopacky}, Q.~M., \& {Stone}, J.~M. 2021, \apj,
  914, 124

\bibitem[{{Burgasser} \& {Splat Development Team}(2017)}]{Burgasser2017}
{Burgasser}, A.~J., \& {Splat Development Team}. 2017, in Astronomical Society
  of India Conference Series, Vol.~14, Astronomical Society of India Conference
  Series, 7--12

\bibitem[{{Burrows} {et~al.}(2001){Burrows}, {Hubbard}, {Lunine}, \&
  {Liebert}}]{Burrows2001}
{Burrows}, A., {Hubbard}, W.~B., {Lunine}, J.~I., \& {Liebert}, J. 2001,
  Reviews of Modern Physics, 73, 719

\bibitem[{{Currie} {et~al.}(2011){Currie}, {Burrows}, {Itoh}, {Matsumura},
  {Fukagawa}, {Apai}, {Madhusudhan}, {Hinz}, {Rodigas}, {Kasper}, {Pyo}, \&
  {Ogino}}]{Currie2011}
{Currie}, T., {Burrows}, A., {Itoh}, Y., {et~al.} 2011, \apj, 729, 128

\bibitem[{{Currie} {et~al.}(2014){Currie}, {Burrows}, {Girard}, {Cloutier},
  {Fukagawa}, {Sorahana}, {Kuchner}, {Kenyon}, {Madhusudhan}, {Itoh},
  {Jayawardhana}, {Matsumura}, \& {Pyo}}]{Currie2014}
{Currie}, T., {Burrows}, A., {Girard}, J.~H., {et~al.} 2014, \apj, 795, 133

\bibitem[{{Cutri} \& {et al.}(2012)}]{Cutri2012}
{Cutri}, R.~M., \& {et al.} 2012, VizieR Online Data Catalog, II/311

\bibitem[{{Cutri} {et~al.}(2003){Cutri}, {Skrutskie}, {van Dyk}, {Beichman},
  {Carpenter}, {Chester}, {Cambresy}, {Evans}, {Fowler}, {Gizis}, {Howard},
  {Huchra}, {Jarrett}, {Kopan}, {Kirkpatrick}, {Light}, {Marsh}, {McCallon},
  {Schneider}, {Stiening}, {Sykes}, {Weinberg}, {Wheaton}, {Wheelock}, \&
  {Zacarias}}]{Cutri2003}
{Cutri}, R.~M., {Skrutskie}, M.~F., {van Dyk}, S., {et~al.} 2003, VizieR Online
  Data Catalog, II/246

\bibitem[{{Doelman} {et~al.}(2020){Doelman}, {Por}, {Ruane}, {Escuti}, \&
  {Snik}}]{Doelman2020}
{Doelman}, D.~S., {Por}, E.~H., {Ruane}, G., {Escuti}, M.~J., \& {Snik}, F.
  2020, \pasp, 132, 045002

\bibitem[{{Doelman} {et~al.}(2017){Doelman}, {Snik}, {Warriner}, \&
  {Escuti}}]{Doelman2017}
{Doelman}, D.~S., {Snik}, F., {Warriner}, N.~Z., \& {Escuti}, M.~J. 2017, in
  Society of Photo-Optical Instrumentation Engineers (SPIE) Conference Series,
  Vol. 10400, Society of Photo-Optical Instrumentation Engineers (SPIE)
  Conference Series, 104000U

\bibitem[{Doelman {et~al.}(2021)Doelman, Snik, Por, Bos, Otten, Kenworthy,
  Haffert, Wilby, Bohn, Sutlieff, Miller, Ouellet, de~Boer, Keller, Escuti,
  Shi, Warriner, Hornburg, Birkby, Males, Morzinski, Close, Codona, Long,
  Schatz, Lumbres, Rodack, Gorkom, Hedglen, Guyon, Lozi, Groff, Chilcote,
  Jovanovic, Thibault, de~Jonge, Allain, Vall\'{e}e, Patel, C\^{o}t\'{e},
  Marois, Hinz, Stone, Skemer, Briesemeister, Boehle, Glauser, Taylor, Baudoz,
  Huby, Absil, Carlomagno, \& Delacroix}]{Doelman:21}
Doelman, D.~S., Snik, F., Por, E.~H., {et~al.} 2021, Appl. Opt., 60, D52

\bibitem[{{Fabrycky} \& {Murray-Clay}(2010)}]{Fabrycky2010}
{Fabrycky}, D.~C., \& {Murray-Clay}, R.~A. 2010, \apj, 710, 1408

\bibitem[{{Finger} {et~al.}(2008){Finger}, {Dorn}, {Eschbaumer}, {Hall},
  {Mehrgan}, {Meyer}, \& {Stegmeier}}]{Finger2008}
{Finger}, G., {Dorn}, R.~J., {Eschbaumer}, S., {et~al.} 2008, in Society of
  Photo-Optical Instrumentation Engineers (SPIE) Conference Series, Vol. 7021,
  High Energy, Optical, and Infrared Detectors for Astronomy III, ed. D.~A.
  {Dorn} \& A.~D. {Holland}, 70210P

\bibitem[{{Gaia Collaboration} {et~al.}(2018){Gaia Collaboration}, {Brown},
  {Vallenari}, {Prusti}, {de Bruijne}, {et~al.}}]{GaiaCollaboration2018}
{Gaia Collaboration}, {Brown}, A.~G.~A., {Vallenari}, A., {et~al.} 2018, \aap,
  616, A1

\bibitem[{{Galicher} {et~al.}(2011){Galicher}, {Marois}, {Macintosh}, {Barman},
  \& {Konopacky}}]{Galicher2011}
{Galicher}, R., {Marois}, C., {Macintosh}, B., {Barman}, T., \& {Konopacky}, Q.
  2011, \apjl, 739, L41

\bibitem[{{Greco} \& {Brandt}(2016)}]{Greco2016}
{Greco}, J.~P., \& {Brandt}, T.~D. 2016, \apj, 833, 134

\bibitem[{{Greenbaum} {et~al.}(2018){Greenbaum}, {Pueyo}, {Ruffio}, {Wang}, {De
  Rosa}, {Aguilar}, {Rameau}, {Barman}, {Marois}, {Marley}, {Konopacky},
  {Rajan}, {Macintosh}, {Ansdell}, {Arriaga}, {Bailey}, {Bulger}, {Burrows},
  {Chilcote}, {Cotten}, {Doyon}, {Duch{\^e}ne}, {Fitzgerald}, {Follette},
  {Gerard}, {Goodsell}, {Graham}, {Hibon}, {Hung}, {Ingraham}, {Kalas},
  {Larkin}, {Maire}, {Marchis}, {Metchev}, {Millar-Blanchaer}, {Nielsen},
  {Norton}, {Oppenheimer}, {Palmer}, {Patience}, {Perrin}, {Poyneer},
  {Rantakyr{\"o}}, {Savransky}, {Schneider}, {Sivaramakrishnan}, {Song},
  {Soummer}, {Thomas}, {Wallace}, {Ward-Duong}, {Wiktorowicz}, \&
  {Wolff}}]{Greenbaum2018}
{Greenbaum}, A.~Z., {Pueyo}, L., {Ruffio}, J.-B., {et~al.} 2018, \aj, 155, 226

\bibitem[{{Hinz} {et~al.}(2018){Hinz}, {Skemer}, {Stone}, {Montoya}, \&
  {Durney}}]{Hinz2018}
{Hinz}, P.~M., {Skemer}, A., {Stone}, J., {Montoya}, O.~M., \& {Durney}, O.
  2018, in Society of Photo-Optical Instrumentation Engineers (SPIE) Conference
  Series, Vol. 10702, Ground-based and Airborne Instrumentation for Astronomy
  VII, ed. C.~J. {Evans}, L.~{Simard}, \& H.~{Takami}, 107023L

\bibitem[{{Hinz} {et~al.}(2016){Hinz}, {Defr{\`e}re}, {Skemer}, {Bailey},
  {Stone}, {Spalding}, {Vaz}, {Pinna}, {Puglisi}, {Esposito}, {Montoya},
  {Downey}, {Leisenring}, {Durney}, {Hoffmann}, {Hill}, {Millan-Gabet},
  {Mennesson}, {Danchi}, {Morzinski}, {Grenz}, {Skrutskie}, \&
  {Ertel}}]{Hinz2016}
{Hinz}, P.~M., {Defr{\`e}re}, D., {Skemer}, A., {et~al.} 2016, in Society of
  Photo-Optical Instrumentation Engineers (SPIE) Conference Series, Vol. 9907,
  Optical and Infrared Interferometry and Imaging V, ed. F.~{Malbet}, M.~J.
  {Creech-Eakman}, \& P.~G. {Tuthill}, 990704

\bibitem[{{H{\o}g} {et~al.}(2000){H{\o}g}, {Fabricius}, {Makarov}, {Urban},
  {Corbin}, {Wycoff}, {Bastian}, {Schwekendiek}, \& {Wicenec}}]{Hog2000}
{H{\o}g}, E., {Fabricius}, C., {Makarov}, V.~V., {et~al.} 2000, \aap, 355, L27

\bibitem[{{Horne}(1986)}]{Horne1986}
{Horne}, K. 1986, \pasp, 98, 609

\bibitem[{{Hubeny} \& {Burrows}(2007)}]{Hubeny2007}
{Hubeny}, I., \& {Burrows}, A. 2007, \apj, 669, 1248

\bibitem[{{Hunter}(2007)}]{matplotlib}
{Hunter}, J.~D. 2007, Computing in Science Engineering, 9, 90

\bibitem[{{Janson} {et~al.}(2010){Janson}, {Bergfors}, {Goto}, {Brandner}, \&
  {Lafreni{\`e}re}}]{Janson2010}
{Janson}, M., {Bergfors}, C., {Goto}, M., {Brandner}, W., \& {Lafreni{\`e}re},
  D. 2010, \apjl, 710, L35

\bibitem[{{Konopacky} {et~al.}(2013){Konopacky}, {Barman}, {Macintosh}, \&
  {Marois}}]{Konopacky2013}
{Konopacky}, Q.~M., {Barman}, T.~S., {Macintosh}, B.~A., \& {Marois}, C. 2013,
  Science, 339, 1398

\bibitem[{{Leisenring} {et~al.}(2012){Leisenring}, {Skrutskie}, {Hinz},
  {Skemer}, {Bailey}, {Eisner}, {Garnavich}, {Hoffmann}, {Jones}, {Kenworthy},
  {Kuzmenko}, {Meyer}, {Nelson}, {Rodigas}, {Wilson}, \&
  {Vaitheeswaran}}]{Leisenring2012}
{Leisenring}, J.~M., {Skrutskie}, M.~F., {Hinz}, P.~M., {et~al.} 2012, in
  Society of Photo-Optical Instrumentation Engineers (SPIE) Conference Series,
  Vol. 8446, Ground-based and Airborne Instrumentation for Astronomy IV, 84464F

\bibitem[{{Marleau} \& {Cumming}(2014)}]{Marleau2014}
{Marleau}, G.~D., \& {Cumming}, A. 2014, \mnras, 437, 1378

\bibitem[{{Marley} {et~al.}(2007){Marley}, {Fortney}, {Hubickyj},
  {Bodenheimer}, \& {Lissauer}}]{Marley2007}
{Marley}, M.~S., {Fortney}, J.~J., {Hubickyj}, O., {Bodenheimer}, P., \&
  {Lissauer}, J.~J. 2007, \apj, 655, 541

\bibitem[{{Marley} {et~al.}(2012){Marley}, {Saumon}, {Cushing}, {Ackerman},
  {Fortney}, \& {Freedman}}]{Marley2012}
{Marley}, M.~S., {Saumon}, D., {Cushing}, M., {et~al.} 2012, \apj, 754, 135

\bibitem[{{Marley} {et~al.}(2010){Marley}, {Saumon}, \&
  {Goldblatt}}]{Marley2010}
{Marley}, M.~S., {Saumon}, D., \& {Goldblatt}, C. 2010, \apjl, 723, L117

\bibitem[{{Marois} {et~al.}(2006){Marois}, {Lafreni{\`e}re}, {Doyon},
  {Macintosh}, \& {Nadeau}}]{Marois2006}
{Marois}, C., {Lafreni{\`e}re}, D., {Doyon}, R., {Macintosh}, B., \& {Nadeau},
  D. 2006, \apj, 641, 556

\bibitem[{{Marois} {et~al.}(2008){Marois}, {Macintosh}, {Barman}, {Zuckerman},
  {Song}, {Patience}, {Lafreni{\`e}re}, \& {Doyon}}]{Marois2008}
{Marois}, C., {Macintosh}, B., {Barman}, T., {et~al.} 2008, Science, 322, 1348

\bibitem[{{Marois} {et~al.}(2010){Marois}, {Zuckerman}, {Konopacky},
  {Macintosh}, \& {Barman}}]{Marois2010}
{Marois}, C., {Zuckerman}, B., {Konopacky}, Q.~M., {Macintosh}, B., \&
  {Barman}, T. 2010, \nat, 468, 1080

\bibitem[{{Molli{\`e}re} {et~al.}(2020){Molli{\`e}re}, {Stolker}, {Lacour},
  {Otten}, {Shangguan}, {Charnay}, {Molyarova}, {Nowak}, {Henning}, {Marleau},
  {Semenov}, {van Dishoeck}, {Eisenhauer}, {Garcia}, {Garcia Lopez}, {Girard},
  {Greenbaum}, {Hinkley}, {Kervella}, {Kreidberg}, {Maire}, {Nasedkin},
  {Pueyo}, {Snellen}, {Vigan}, {Wang}, {de Zeeuw}, \& {Zurlo}}]{Molliere2020}
{Molli{\`e}re}, P., {Stolker}, T., {Lacour}, S., {et~al.} 2020, \aap, 640, A131

\bibitem[{{Oppenheimer} {et~al.}(2013){Oppenheimer}, {Baranec}, {Beichman},
  {Brenner}, {Burruss}, {Cady}, {Crepp}, {Dekany}, {Fergus}, {Hale},
  {Hillenbrand}, {Hinkley}, {Hogg}, {King}, {Ligon}, {Lockhart}, {Nilsson},
  {Parry}, {Pueyo}, {Rice}, {Roberts}, {Roberts}, {Shao}, {Sivaramakrishnan},
  {Soummer}, {Truong}, {Vasisht}, {Veicht}, {Vescelus}, {Wallace}, {Zhai}, \&
  {Zimmerman}}]{Oppenheimer2013}
{Oppenheimer}, B.~R., {Baranec}, C., {Beichman}, C., {et~al.} 2013, \apj, 768,
  24

\bibitem[{{Otten} {et~al.}(2017){Otten}, {Snik}, {Kenworthy}, {Keller},
  {Males}, {Morzinski}, {Close}, {Codona}, {Hinz}, {Hornburg}, {Brickson}, \&
  {Escuti}}]{Otten2017}
{Otten}, G. P.~P.~L., {Snik}, F., {Kenworthy}, M.~A., {et~al.} 2017, \apj, 834,
  175

\bibitem[{{Patience} {et~al.}(2010){Patience}, {King}, {de Rosa}, \&
  {Marois}}]{Patience2010}
{Patience}, J., {King}, R.~R., {de Rosa}, R.~J., \& {Marois}, C. 2010, \aap,
  517, A76

\bibitem[{Possolo {et~al.}(2019)Possolo, Merkatas, \& Bodnar}]{Possolo2019}
Possolo, A., Merkatas, C., \& Bodnar, O. 2019, Metrologia, 56, 045009

\bibitem[{{Rajan} {et~al.}(2017){Rajan}, {Rameau}, {De Rosa}, {Marley},
  {Graham}, {Macintosh}, {Marois}, {Morley}, {Patience}, {Pueyo}, {Saumon},
  {Ward-Duong}, {Ammons}, {Arriaga}, {Bailey}, {Barman}, {Bulger}, {Burrows},
  {Chilcote}, {Cotten}, {Czekala}, {Doyon}, {Duch{\^e}ne}, {Esposito},
  {Fitzgerald}, {Follette}, {Fortney}, {Goodsell}, {Greenbaum}, {Hibon},
  {Hung}, {Ingraham}, {Johnson-Groh}, {Kalas}, {Konopacky}, {Lafreni{\`e}re},
  {Larkin}, {Maire}, {Marchis}, {Metchev}, {Millar-Blanchaer}, {Morzinski},
  {Nielsen}, {Oppenheimer}, {Palmer}, {Patel}, {Perrin}, {Poyneer},
  {Rantakyr{\"o}}, {Ruffio}, {Savransky}, {Schneider}, {Sivaramakrishnan},
  {Song}, {Soummer}, {Thomas}, {Vasisht}, {Wallace}, {Wang}, {Wiktorowicz}, \&
  {Wolff}}]{Rajan2017}
{Rajan}, A., {Rameau}, J., {De Rosa}, R.~J., {et~al.} 2017, \aj, 154, 10

\bibitem[{{Rodrigo} \& {Solano}(2020)}]{Rodrigo2020}
{Rodrigo}, C., \& {Solano}, E. 2020, in XIV.0 Scientific Meeting (virtual) of
  the Spanish Astronomical Society, 182

\bibitem[{{Rodrigo} {et~al.}(2012){Rodrigo}, {Solano}, \& {Bayo}}]{Rodrigo2012}
{Rodrigo}, C., {Solano}, E., \& {Bayo}, A. 2012, {SVO Filter Profile Service
  Version 1.0}, IVOA Working Draft 15 October 2012, , ,
  doi:10.5479/ADS/bib/2012ivoa.rept.1015R

\bibitem[{{Saumon} \& {Marley}(2008)}]{Saumon2008}
{Saumon}, D., \& {Marley}, M.~S. 2008, \apj, 689, 1327

\bibitem[{{Skemer} {et~al.}(2018{\natexlab{a}}){Skemer}, {Hinz}, {Stone},
  {Skrutskie}, {Woodward}, {Leisenring}, \& {Briesemeister}}]{Skemer2018}
{Skemer}, A.~J., {Hinz}, P., {Stone}, J., {et~al.} 2018{\natexlab{a}}, in
  Society of Photo-Optical Instrumentation Engineers (SPIE) Conference Series,
  Vol. 10702, Ground-based and Airborne Instrumentation for Astronomy VII, ed.
  C.~J. {Evans}, L.~{Simard}, \& H.~{Takami}, 107020C

\bibitem[{{Skemer} {et~al.}(2012){Skemer}, {Hinz}, {Esposito}, {Burrows},
  {Leisenring}, {Skrutskie}, {Desidera}, {Mesa}, {Arcidiacono}, {Mannucci},
  {Rodigas}, {Close}, {McCarthy}, {Kulesa}, {Agapito}, {Apai}, {Argomedo},
  {Bailey}, {Boutsia}, {Briguglio}, {Brusa}, {Busoni}, {Claudi}, {Eisner},
  {Fini}, {Follette}, {Garnavich}, {Gratton}, {Guerra}, {Hill}, {Hoffmann},
  {Jones}, {Krejny}, {Males}, {Masciadri}, {Meyer}, {Miller}, {Morzinski},
  {Nelson}, {Pinna}, {Puglisi}, {Quanz}, {Quiros-Pacheco}, {Riccardi},
  {Stefanini}, {Vaitheeswaran}, {Wilson}, \& {Xompero}}]{Skemer2012}
{Skemer}, A.~J., {Hinz}, P.~M., {Esposito}, S., {et~al.} 2012, \apj, 753, 14

\bibitem[{{Skemer} {et~al.}(2014){Skemer}, {Marley}, {Hinz}, {Morzinski},
  {Skrutskie}, {Leisenring}, {Close}, {Saumon}, {Bailey}, {Briguglio},
  {Defrere}, {Esposito}, {Follette}, {Hill}, {Males}, {Puglisi}, {Rodigas}, \&
  {Xompero}}]{Skemer2014}
{Skemer}, A.~J., {Marley}, M.~S., {Hinz}, P.~M., {et~al.} 2014, \apj, 792, 17

\bibitem[{{Skemer} {et~al.}(2015){Skemer}, {Hinz}, {Montoya}, {Skrutskie},
  {Leisenring}, {Durney}, {Woodward}, {Wilson}, {Nelson}, {Bailey}, {Defrere},
  \& {Stone}}]{Skemer2015}
{Skemer}, A.~J., {Hinz}, P., {Montoya}, M., {et~al.} 2015, in Society of
  Photo-Optical Instrumentation Engineers (SPIE) Conference Series, Vol. 9605,
  Techniques and Instrumentation for Detection of Exoplanets VII, ed.
  S.~{Shaklan}, 96051D

\bibitem[{{Skemer} {et~al.}(2016){Skemer}, {Morley}, {Zimmerman}, {Skrutskie},
  {Leisenring}, {Buenzli}, {Bonnefoy}, {Bailey}, {Hinz}, {Defr{\'e}re},
  {Esposito}, {Apai}, {Biller}, {Brandner}, {Close}, {Crepp}, {De Rosa},
  {Desidera}, {Eisner}, {Fortney}, {Freedman}, {Henning}, {Hofmann},
  {Kopytova}, {Lupu}, {Maire}, {Males}, {Marley}, {Morzinski}, {Oza},
  {Patience}, {Rajan}, {Rieke}, {Schertl}, {Schlieder}, {Stone}, {Su}, {Vaz},
  {Visscher}, {Ward-Duong}, {Weigelt}, \& {Woodward}}]{Skemer2016}
{Skemer}, A.~J., {Morley}, C.~V., {Zimmerman}, N.~T., {et~al.} 2016, \apj, 817,
  166

\bibitem[{{Skemer} {et~al.}(2018{\natexlab{b}}){Skemer}, {Stelter}, {Mawet},
  {Fitzgerald}, {Mazin}, {Guyon}, {Marois}, {Briesemeister}, {Brandt},
  {Chilcote}, {Delorme}, {Jovanovic}, {Lu}, {Millar-Blanchaer}, {Wallace},
  {Vasisht}, {Roberts}, \& {Wang}}]{Skemer2018b}
{Skemer}, A.~J., {Stelter}, D., {Mawet}, D., {et~al.} 2018{\natexlab{b}}, in
  Society of Photo-Optical Instrumentation Engineers (SPIE) Conference Series,
  Vol. 10702, Ground-based and Airborne Instrumentation for Astronomy VII, ed.
  C.~J. {Evans}, L.~{Simard}, \& H.~{Takami}, 10702A5

\bibitem[{{Skrutskie} {et~al.}(2010){Skrutskie}, {Jones}, {Hinz}, {Garnavich},
  {Wilson}, {Nelson}, {Solheid}, {Durney}, {Hoffmann}, {Vaitheeswaran},
  {McMahon}, {Leisenring}, \& {Wong}}]{Skrutskie2010}
{Skrutskie}, M.~F., {Jones}, T., {Hinz}, P., {et~al.} 2010, in Society of
  Photo-Optical Instrumentation Engineers (SPIE) Conference Series, Vol. 7735,
  Ground-based and Airborne Instrumentation for Astronomy III, 77353H

\bibitem[{{Snik} {et~al.}(2012){Snik}, {Otten}, {Kenworthy}, {Miskiewicz},
  {Escuti}, {Packham}, \& {Codona}}]{Snik2012}
{Snik}, F., {Otten}, G., {Kenworthy}, M., {et~al.} 2012, in Society of
  Photo-Optical Instrumentation Engineers (SPIE) Conference Series, Vol. 8450,
  Modern Technologies in Space- and Ground-based Telescopes and Instrumentation
  II, ed. R.~{Navarro}, C.~R. {Cunningham}, \& E.~{Prieto}, 84500M

\bibitem[{{Stephens} {et~al.}(2009){Stephens}, {Leggett}, {Cushing}, {Marley},
  {Saumon}, {Geballe}, {Golimowski}, {Fan}, \& {Noll}}]{Stephens2009}
{Stephens}, D.~C., {Leggett}, S.~K., {Cushing}, M.~C., {et~al.} 2009, \apj,
  702, 154

\bibitem[{{Stolker} {et~al.}(2019){Stolker}, {Bonse}, {Quanz}, {Amara},
  {Cugno}, {Bohn}, \& {Boehle}}]{Stolker2019}
{Stolker}, T., {Bonse}, M.~J., {Quanz}, S.~P., {et~al.} 2019, \aap, 621, A59

\bibitem[{{Stone} {et~al.}(2018){Stone}, {Skemer}, {Hinz}, {Briesemeister},
  {Barman}, {Woodward}, {Skrutskie}, \& {Leisenring}}]{Stone2018}
{Stone}, J.~M., {Skemer}, A.~J., {Hinz}, P., {et~al.} 2018, in Society of
  Photo-Optical Instrumentation Engineers (SPIE) Conference Series, Vol. 10702,
  Ground-based and Airborne Instrumentation for Astronomy VII, ed. C.~J.
  {Evans}, L.~{Simard}, \& H.~{Takami}, 107023F

\bibitem[{{Stone} {et~al.}(2016){Stone}, {Eisner}, {Skemer}, {Morzinski},
  {Close}, {Males}, {Rodigas}, {Hinz}, \& {Puglisi}}]{Stone2016}
{Stone}, J.~M., {Eisner}, J., {Skemer}, A., {et~al.} 2016, \apj, 829, 39

\bibitem[{{Virtanen} {et~al.}(2020){Virtanen}, {Gommers}, {Oliphant},
  {et~al.}}]{2020SciPy-NMeth}
{Virtanen}, P., {Gommers}, R., {Oliphant}, T.~E., {et~al.} 2020, Nature
  Methods, doi:https://doi.org/10.1038/s41592-019-0686-2

\bibitem[{{Wagner} {et~al.}(2020){Wagner}, {Stone}, {Dong}, {Ertel}, {Apai},
  {Doelman}, {Bohn}, {Najita}, {Brittain}, {Kenworthy}, {Keppler}, {Webster},
  {Mailhot}, \& {Snik}}]{Wagner2020}
{Wagner}, K., {Stone}, J., {Dong}, R., {et~al.} 2020, \aj, 159, 252

\bibitem[{{Wang} {et~al.}(2018){Wang}, {Mawet}, {Fortney}, {Hood}, {Morley}, \&
  {Benneke}}]{Wang2018}
{Wang}, J., {Mawet}, D., {Fortney}, J.~J., {et~al.} 2018, \aj, 156, 272

\bibitem[{{Wang} {et~al.}(2020){Wang}, {Wang}, {Ma}, {Chilcote}, {Ertel},
  {Guyon}, {Ilyin}, {Jovanovic}, {Kalas}, {Lozi}, {Macintosh}, {Strassmeier},
  \& {Stone}}]{Wang2020}
{Wang}, J., {Wang}, J.~J., {Ma}, B., {et~al.} 2020, \aj, 160, 150

\bibitem[{{Witte} {et~al.}(2011){Witte}, {Helling}, {Barman}, {Heidrich}, \&
  {Hauschildt}}]{Witte2011}
{Witte}, S., {Helling}, C., {Barman}, T., {Heidrich}, N., \& {Hauschildt},
  P.~H. 2011, \aap, 529, A44

\bibitem[{{Zurlo} {et~al.}(2016){Zurlo}, {Vigan}, {Galicher}, {Maire}, {Mesa},
  {Gratton}, {Chauvin}, {Kasper}, {Moutou}, {Bonnefoy}, {Desidera}, {Abe},
  {Apai}, {Baruffolo}, {Baudoz}, {Baudrand}, {Beuzit}, {Blancard},
  {Boccaletti}, {Cantalloube}, {Carle}, {Cascone}, {Charton}, {Claudi},
  {Costille}, {de Caprio}, {Dohlen}, {Dominik}, {Fantinel}, {Feautrier},
  {Feldt}, {Fusco}, {Gigan}, {Girard}, {Gisler}, {Gluck}, {Gry}, {Henning},
  {Hugot}, {Janson}, {Jaquet}, {Lagrange}, {Langlois}, {Llored}, {Madec},
  {Magnard}, {Martinez}, {Maurel}, {Mawet}, {Meyer}, {Milli},
  {Moeller-Nilsson}, {Mouillet}, {Orign{\'e}}, {Pavlov}, {Petit}, {Puget},
  {Quanz}, {Rabou}, {Ramos}, {Rousset}, {Roux}, {Salasnich}, {Salter},
  {Sauvage}, {Schmid}, {Soenke}, {Stadler}, {Suarez}, {Turatto}, {Udry},
  {Vakili}, {Wahhaj}, {Wildi}, \& {Antichi}}]{Zurlo2016}
{Zurlo}, A., {Vigan}, A., {Galicher}, R., {et~al.} 2016, \aap, 587, A57

\end{thebibliography}
\bibliographystyle{aasjournal}
\end{document}